\documentclass[fleqn,usenatbib]{mnras}

\usepackage{newtxtext,newtxmath}

\usepackage[T1]{fontenc}

\DeclareRobustCommand{\VAN}[3]{#2}
\let\VANthebibliography\thebibliography
\def\thebibliography{\DeclareRobustCommand{\VAN}[3]{##3}\VANthebibliography}

\usepackage{graphicx}	
\usepackage{amsmath}	
\usepackage{siunitx}
\usepackage{xcolor}

\newcommand{\rone}{FRB~20121102A}
\newcommand{\rthree}{FRB~20180916B}
\newcommand{\rsixseven}{FRB~20201124A}

\defcitealias{pleunisetal}{P21}

\newcommand{\pcpercc}{pc~cm$^{-3}$}
\newcommand{\radpermsq}{rad~m$^{-2}$}
\newcommand{\mhzperms}{MHz~ms$^{-1}$}

\title[\rthree: long-term monitoring with LOFAR]{Propagation effects at low frequencies seen in the LOFAR long-term monitoring of the periodically active \rthree}

\author[A.~Gopinath et al.]{A.~Gopinath,$^{1}$\thanks{E-mail: a.gopinath@uva.nl}
C.~G.~Bassa,$^{2}$
Z.~Pleunis,$^{3}$
J.~W.~T.~Hessels,$^{1,2}$
P.~Chawla,$^{1}$
E.~F.~Keane,$^{6}$
V.~Kondratiev,$^{2}$ \newauthor
D.~Michilli$^{4,5}$
and
K.~Nimmo$^{5}$
\\
$^{1}$Anton Pannekoek Institute for Astronomy, University of Amsterdam, Science Park 904, 1098 XH Amsterdam, The Netherlands \\
$^{2}$ASTRON, Netherlands Institute for Radio Astronomy, Oude Hoogeveensedijk 4, 7991 PD Dwingeloo, The Netherlands \\
$^{3}$Dunlap Institute for Astronomy \& Astrophysics, University of Toronto, 50 St. George Street, Toronto, ON M5S 3H4, Canada \\
$^{4}$Department of Physics, Massachusetts Institute of Technology, 77 Massachusetts Ave, Cambridge, MA 02139, USA \\
$^{5}$MIT Kavli Institute for Astrophysics and Space Research, Massachusetts Institute of Technology, 77 Massachusetts Ave, Cambridge, MA 02139, USA \\
$^{6}$School of Physics, Trinity College Dublin, College Green, Dublin 2, D02 PN40, Ireland
}

\date{Accepted XXX. Received YYY; in original form ZZZ}

\pubyear{2023}

\begin{document}
\label{firstpage}
\pagerange{\pageref{firstpage}--\pageref{lastpage}}
\maketitle

\begin{abstract}
LOFAR (LOw Frequency ARray) has previously detected bursts from the periodically active, repeating fast radio burst (FRB) source \rthree\ down to unprecedentedly low radio frequencies of 110\,MHz. Here we present 11 new bursts in 223 more hours of continued monitoring of \rthree\ in the 110$-$188\,MHz  band with LOFAR. We place new constraints on the source's activity window $w =4.3^{+0.7}_{-0.2}$\,day and phase centre $\phi_{\mathrm{c}}^{\mathrm{LOFAR}} = 0.67^{+0.03}_{-0.02}$ in its 16.33-day activity cycle, strengthening evidence for its frequency-dependent activity cycle. Propagation effects like Faraday rotation and scattering are especially pronounced at low frequencies and constrain properties of FRB 20180916B’s local environment. We track variations in scattering and time-frequency drift rates, and find no evidence for trends in time or activity phase. Faraday rotation measure (RM) variations seen between June 2021 and August 2022 show a fractional change $>$50\% with hints of flattening of the gradient of the previously reported secular trend seen at 600\,MHz. The frequency-dependent window of activity at LOFAR appears stable despite the significant changes in RM, leading us to deduce that these two effects have different causes. Depolarization of and within individual bursts towards lower radio frequencies is quantified using LOFAR’s large fractional bandwidth, with some bursts showing no detectable polarization. However, the degree of depolarization seems uncorrelated to the scattering timescales, allowing us to evaluate different depolarization models. We discuss these results in the context of models that invoke rotation, precession, or binary orbital motion to explain the periodic activity of \rthree.
\end{abstract}




\section{Introduction}\label{sec:intro}

Fast radio bursts (FRBs; see \citealt{petroff_2022_aarv} for a review) are astrophysical transients that produce millisecond-duration coherent radio emission. Their characteristic large dispersive delays are the result of their extragalactic distances, which have been confirmed in some cases by a robust association with a host galaxy \citep[e.g.,][]{chatterjee_2017_natur, bannister_2019_sci, marcote_2020_natur, kirsten_2022_natur, Niu_2022_Natur, ravi_2022_mnras}. They are many orders-of-magnitude more luminous than the individual pulses from Galactic pulsars, including even the giant pulses from the young Crab pulsar. Since the serendipitous discovery of the `Lorimer burst' \citep[FRB~20010724A;][]{lorimer_2007_sci}, over 600 FRB sources have been catalogued, most by the Canadian Hydrogen Intensity Mapping Experiment's FRB detection system \citep[CHIME/FRB;][]{chime_2021_apjs, chime_2023_arXiv}. In this sample, most have been one-off events (apparent non-repeating FRBs); $\approx 4\%$ of FRBs have been seen to repeat \citep{spitler_2016_natur,pleunis_2021_apj, chime_2023_arXiv} --- \rthree\ with a periodic activity cycle of $16.33 \pm 0.12$\,day \citep{chime_2020_natur_periodicactivity, pleunis_2021_apj}, and \rone\ with a possible $\approx160$-day periodicity \citep{rajwade_2020_mnras, cruces_2021_mnras}. It has been argued that the apparent one-off events are less active repeating sources that have simply not been observed to repeat yet \citep{caleb_2019_mnras_484, ravi_2019_natas,james_2020_mnras,chime_2023_arXiv}. However, burst morphological and spectral differences between repeating and apparently non-repeating FRBs point to the possible existence of at least two distinct populations of FRBs \citep{hessels_2019_apjl,pleunis_2021_apj,chime_2023_arXiv}. Alternatively, this dichotomy may indicate a different burst mechanism in the same type of progenitor source \citep{Kirsten_2023_arXiv}, or may be a consequence of beaming direction \citep[at least in the case of burst duration;][]{Connor_2020_MNRAS}.

We can further explore the origins of FRBs through studies of their host galaxy environments, changing burst activity \citep{li_2021_natur, Hewitt_2022_MNRAS, Nimmo_2023_MNRAS}, and evolution of burst properties such as dispersion measure (DM) and Faraday rotation measure \citep[RM;][]{michilli_2018_natur,hilmarsson_2021_apjl}. This can help establish differences, if any, in the hosts \citep{bhandari_2022_aj} and local environments of FRBs \citep{kirsten_2022_natur}.

\rthree\ has a $16.33\pm0.12$-day period in its activity, though no strict periodicity has been seen between burst arrival times \citep{chime_2020_natur_periodicactivity, pleunis_2021_apj}.
The surprising discovery of this periodic activity led to models in which the source may be: i) a highly magnetised neutron star (NS) in an interacting binary system \citep{ioka_2020_apjl, lyutikov_2020_apjl, Barkov_2022_arXiv}; ii) an extremely slowly rotating \citep{beniamini_2020_mnras_496}, and presumably old ($1 - 10$\,kyr), magnetar; or, iii) a precessing \citep{levin_2020_apjl, li_2021_apjl}, presumably young ($< 1$\,kyr), magnetar.

\rthree\ was localized to the vicinity of a star-forming region in a spiral host galaxy  \citep{marcote_2020_natur,tendulkar_2021_apjl}. At a luminosity distance of $\sim 149$\,Mpc ($z = 0.0337 \pm 0.0002$), it is offset by about $250$\,pc from the peak brightness of the nearest star-forming knot in its host galaxy. Assuming that it was born in this region and has subsequently moved away from it, the inferred age of $\sim10^{5-7}$\,yr is inconsistent with that of a young magnetar, and instead led \citet{tendulkar_2021_apjl} to suggest a high-mass X-ray or gamma-ray binary progenitor model (though an OB runaway scenario cannot be ruled out). The firmly established activity period \citep{chime_2020_natur_periodicactivity} aids multi-frequency follow-up observations of the source because it provides some predictability to use in the planning of observations. As a result, \rthree\ has been detected at over 4 octaves in radio frequency, from $110$\,MHz \citep[][hereafter P21]{pleunisetal} up to $5400$\,MHz \citep{chawla_2020_apjl, pilia_2020_apjl, Marthi_2020_MNRAS, marcote_2020_natur, nimmo_2021_natas, pastormarazuela_2021_natur, Sand_2022_ApJ, Bethapudi_2022_arXiv}. The Low-Frequency Array (LOFAR) High Band Antenna (HBA) detection of 18 bursts \citepalias{pleunisetal} strongly constrains free-free absorption in the local environment of the source. These detections also demonstrated that \rthree's periodic activity is systematically delayed towards lower frequencies, in addition to detecting subtle, but measurable, RM variations of $2-4$\,\radpermsq\ \citepalias[$2-3$\% fractional variations]{pleunisetal}. 

The LOFAR telescope \citep{vanHaarlem_2013_A&A} operates at the lowest radio frequencies ($\nu$) detectable from the Earth's surface (just above the ionospheric cut-off). At these low frequencies, propagation effects such as scattering ($\propto \nu^{-4}$), dispersive delay ($\propto \nu^{-2}$), and Faraday rotation ($\propto \nu^{-2}$) are much more pronounced. The temporal broadening due to increased scattering can lead to a reduction in the signal-to-noise ratio (S/N) of the bursts, making detection more challenging. Conversely, this also means that these effects can be measured and tracked for changes at very high precision with LOFAR -- which offers $\sim0.1$\,\radpermsq\ precision for tracking subtle, but still significant\footnote{Note that, at this level of precision, ionospheric RM variations must also be considered \citep{Sotomayor-Beltran_2013_A&A}.}, RM variations of a few \radpermsq\ \citepalias{pleunisetal}.  

The spectro-temporal properties of the bursts give insight into the local environment and emission mechanism. Like most repeaters, \rthree's bursts often show a $\sim10-30$\% fractional bandwidth of emission and sub-burst drifting structure, argued to be caused by a radius-to-frequency mapping in the magnetosphere of a NS \citep{hessels_2019_apjl, Wang_2019_ApJL, lyutikov_2020_apj}. The bursts have been observed to be $\sim100$\% linearly polarized at high frequencies \citep{nimmo_2021_natas}, with no detectable circular polarization, and depolarizing only at the lowest frequencies \citepalias{pleunisetal}. This depolarization has been observed for other repeating FRBs as well \citep{Feng_2022_Sci}, including the other possibly periodically repeating \rone\ \citep{Plavin_2022_MNRAS}. In addition, the flat polarization position angles (PPAs) that remain consistent even between bursts \citep{nimmo_2021_natas,michilli_2018_natur} disfavour rotational and precession models that expect PPA variations as a function of both the rotational and precession phases. 

Recently, \citet{Mckinven_2023_ApJ_R3} observed a secular change in the RM of \rthree\ using CHIME/FRB, further demonstrating a progenitor residing in a dynamic magnetized environment. This large fractional RM change (of $> 40$\% over a 9-month duration) came after measuring roughly stable RMs with comparatively small variations ($< 10$\%) for nearly three years since its discovery in 2018 (\citealt{chime_2019_apjl}; \citetalias{pleunisetal}). 

In this paper, we present results from a monitoring campaign of \rthree\ with LOFAR at $110-188$\,MHz spanning 78 activity cycles spread across nearly three years. We examine whether the chromatic shift in activity window (\citetalias{pleunisetal}; \citealt{pastormarazuela_2021_natur}; \citealt{Bethapudi_2022_arXiv}) remains constant in time, look for changes in burst rate, and compare our bursts to those detected by CHIME/FRB at $400-800$\,MHz in the same time period. We also track scattering, Faraday rotation measure, and depolarization over multi-year timescales. Section~\ref{sec:obs} describes the observational setup and the burst search; Section~\ref{sec:results} presents the analysis and results of our observations, whose implications are discussed in Section~\ref{sec:disc}.  


\section{Observations and burst search}
\label{sec:obs}

\begin{figure*}
  \centering
  \includegraphics[width=1.0\textwidth]{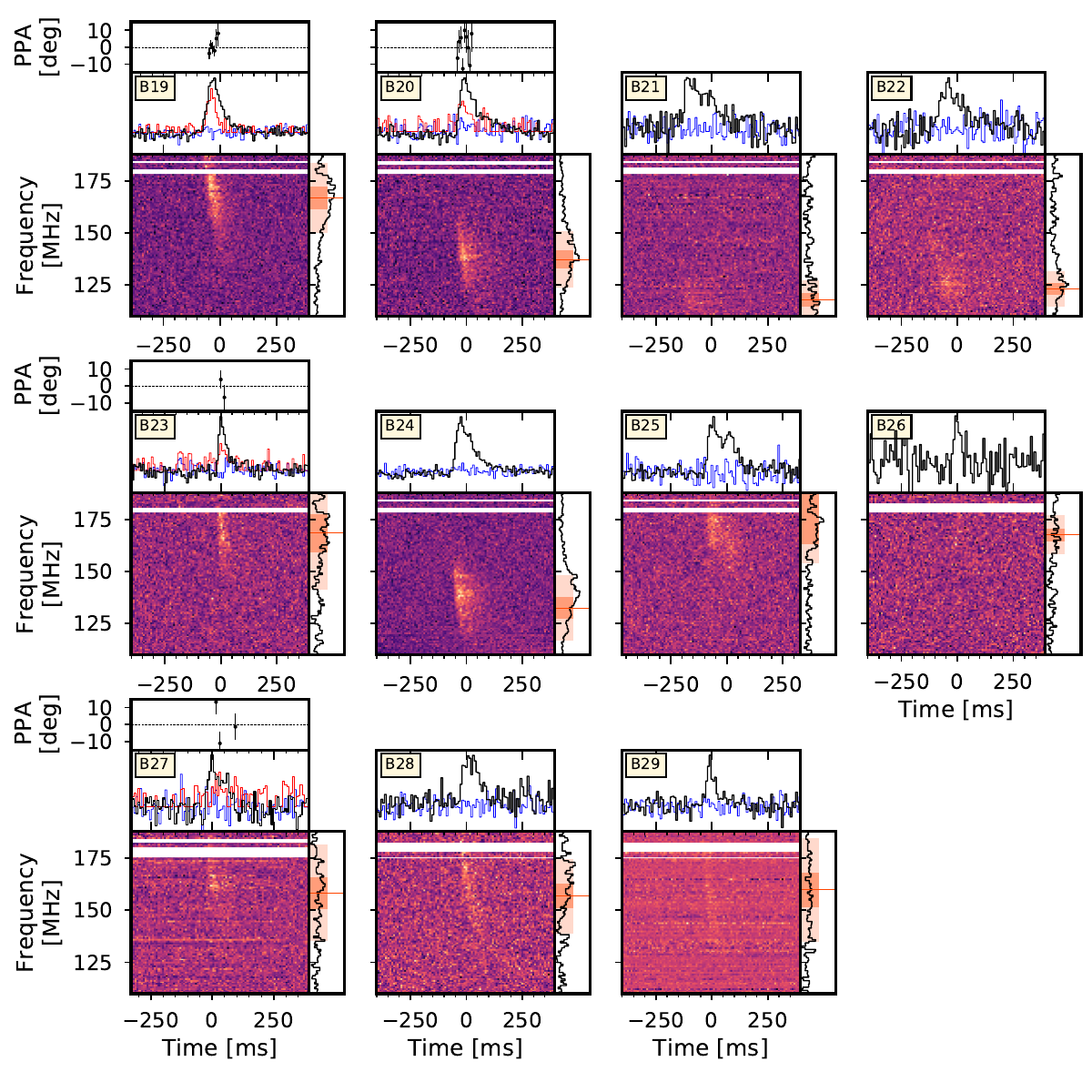}
  \caption{Dynamic spectra (bottom panels of sub-figures), total intensity and polarimetric profiles (mid panels of sub-figures) and PPAs (top panels of sub-figures, for bursts with detected linear polarization) --- for the 11 new bursts (B19$-$B29) detected from \rthree\ during the LOFAR campaign reported here. The PPAs for each burst have been rotated by an arbitrary angle such that they are centered around 0\,deg). All bursts are dedispersed to a DM $=$ 348.772\,\pcpercc, the best-fit DM from \citet{nimmo_2021_natas} who used burst micro-structure to determine this value. See Section~\ref{ssec:scatteringdrift} in the main text for a detailed justification of why we used this single DM value. The spectra have a time and frequency resolution of 7.864\,ms and 0.781\,MHz, respectively. All the time series profiles --- total-intensity Stokes~I (in black), linear polarization L (in red), and circular polarization Stokes~V (in blue) --- are obtained by summing only over the part of the band containing the burst. The white horizontal sections of the dynamic spectra represent masked RFI bands. The central frequency and 3-$\sigma$ frequency extent of the bursts (upon fitting a Gaussian profile to the burst spectrum) are shown in the panels on the right of the dynamic spectrum in each sub-figure.}
  \label{fig:burstfamily}
\end{figure*}

\begin{figure*}
  \centering
  \footnotesize
  \includegraphics[width=1.0\textwidth]{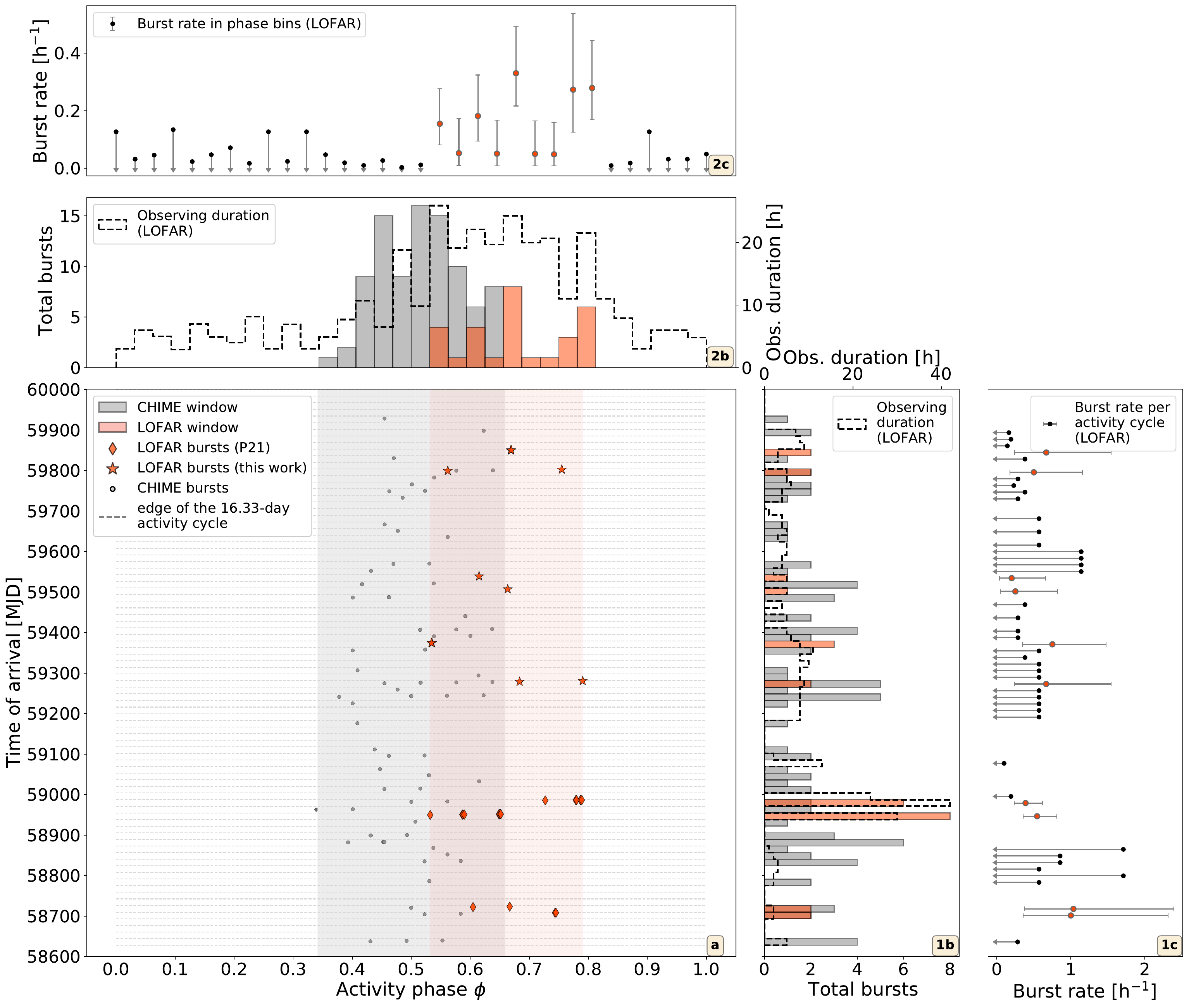}
  \caption[thisiscaption]{Summary of LOFAR observations and detections from \rthree, from monitoring the source over a period of $\sim 3.5$\,yr between 2019 June 06 and 2022 November 17. Panel (a): Barycentric burst arrival times of all bursts from \rthree\ detected with LOFAR and CHIME/FRB\footnotemark[2], versus activity phase. Phases are calculated for a 16.33-day activity period, with the reference MJD~58369.40 set to $\phi = 0.0$, that puts the mean of the phases of CHIME/FRB bursts in \citetalias{pleunisetal} at $\phi = 0.5$. The LOFAR activity window has been shaded in orange (0.53 $\leq \phi \leq$ 0.79), and the CHIME/FRB activity window in gray (0.34 $\leq \phi \leq$ 0.64). Panels (1b) and (2b): Total bursts and exposure at LOFAR, versus time and activity phase, respectively. Panel (1c): burst rate per hour at LOFAR, versus time, per activity cycle. Panel (2c): burst rate per hour at LOFAR, versus activity phase, calculated over all cycles we have exposures for. Detection burst rates are in orange and upper limits are in black. In both panels (1c) and (2c), we plot the 1-$\sigma$ Poissonian errors on the burst rates.}
  \label{fig:activity}
\end{figure*} 

LOFAR has observed \rthree\ between 2019 June 6 and 2022 November 17, with observations of the source during 51 of the 78 16.33-day activity cycles in this time. Results of the observations between 2019 June 6 to 2020 August 6 (13 activity cycles), with 18 bursts detected, were reported in \citetalias{pleunisetal} \citep[half of the same LOFAR bursts were reported in][]{pastormarazuela_2021_natur}. 

For the work presented here, we followed two observing strategies, totaling to 223\,h of exposure on source, spread across 38 activity cycles: (1) Observations concentrated during the expected activity window as reported in \citetalias{pleunisetal}, in order to maximize our chances of detecting high-S/N bursts to measure scattering timescales and RM variations. We followed this strategy between 2021 June to 2021 November, and 2022 June to 2022 November (89\,h). (2) Observations distributed nearly uniformly throughout the 16.33-day activity cycle, in order to establish a well-bounded activity window and duty cycle. We followed this strategy between 2020 December to 2021 May, and 2021 December to 2022 May (134\,h). 

For each observation, signals from the HBAs of up to 24 LOFAR Core stations were coherently added in the \textsc{COBALT2.0} beamformer \citep{Broekema_2018_A&C} to form a single tied-array beam pointing towards the source. We used a pointing direction of $\alpha_\mathrm{J2000}=01^\mathrm{h}58^\mathrm{m}00\fs7502$, $\delta_\mathrm{J2000}=+65\degr43\arcmin00\farcs3152$ (2.3-mas uncertainty), from \citet{marcote_2020_natur}, for all the observations. The $\sim3\arcmin$ Full-Width-Half-Maximum (FWHM) of the tied-array beam covered the uncertainty region of the FRB localization. Nyquist sampled complex voltage data for two linear polarizations were recorded for 400 subbands of $195.3125$\,kHz each, providing $78.125$\,MHz of bandwidth between $110-188$\,MHz, with typical observation durations of 1\,h.

The complex voltage data were analysed using the pipeline described in more detail in
\citetalias{pleunisetal}; we briefly repeat the main aspects here. We channelized the complex
voltage data to a frequency resolution of $3.05$\,kHz (64 channels per subband), after which the two orthogonal polarizations were squared and summed to form Stokes I total intensity data in filterbank format \citep{lorimer_2011_ascl}. These data were averaged to a time resolution of $983.04$\,$\upmu$s (after averaging by a factor of 3), and subsequently averaged in frequency by a factor of 16 using \texttt{digifil} \citep{vanstraten_2011_pasa}, and using incoherent dedispersion to a DM of $348.772$\,\pcpercc, the best-fit DM from \citet{nimmo_2021_natas}. The temporal smearing due to incoherent dedispersion in a single channel at the given DM is between $1.3$\,ms (at 188\,MHz) and $6.7$\,ms (at 110\,MHz). Given the expected DM smearing, intrinsic burst widths (FWHM $>30$\,ms), and scattering timescales ($\sim 50$\,ms) measured for this source in \citetalias{pleunisetal}, we decide that a time resolution of $983.04$\,$\upmu$s is sufficient to detect bursts from \rthree. This approach to creating the dedispersed filterbank files is identical to that used in \citetalias{pleunisetal}, such that the results can be compared. Note that the data were coherently dedispersed for subsequent measurement and analysis of burst properties, as presented in Section~\ref{ssec:pol}.

We used \texttt{rfifind} from the \textsc{PRESTO} software suite \citep{ransom_2001_phdt}
to identify radio frequency interference (RFI); we replaced the values of these frequency channels and time integrations with random noise with the appropriate local statistics. From these RFI cleaned files, dedispersed time-series filterbanks were generated using the graphics processing unit (GPU)-accelerated \textsc{Dedisp}
library \citep{barsdell_2012_mnras} for $\mathrm{DM}$s ranging from $-20$ to
$+20$\,\pcpercc\ around the nominal DM of \rthree, in steps of
0.1\,\pcpercc. 

\footnotetext[2]{The plotted CHIME/FRB arrival times are based on what is reported at \url{https://www.chime-frb.ca/repeaters/FRB20180916B}.}

\rthree\ and other repeating FRBs often show bursts with fractional bandwidths of $\sim10-30$\% \citep{gourdji_2019_apjl,pleunis_2021_apj, kumar_2021_mnras}. To increase the sensitivity to such bursts, dedispersed time-series were also created for three overlapping halves of the band ($110-149$, $130-169$, and $149-188$\,MHz), and seven overlapping quarters of the band ($110-129$, $120-139$\,MHz, etc.). 

\setcounter{footnote}{2}

Candidate bursts were generated by cross-correlating the dedispersed time-series with top-hat functions with widths up to $150$\,ms, using a GPU-accelerated version of \textsc{PRESTO}'s \texttt{single\_pulse\_search.py}\footnote{\url{https://github.com/cbassa/sp_search_gpu}}. The diagnostic plots from \texttt{single\_pulse\_search.py} showing the S/N of bursts versus DM and time were inspected by eye for bursts, as a fail-safe. The events identified by \texttt{single\_pulse\_search.py} at separate DMs were grouped using \texttt{SinglePulseSearcher}\footnote{\url{https://github.com/danielemichilli/SpS}} \citep{michilli_2018_mnras} to eliminate redundant burst candidates obtained from dedispersing the same burst at slightly different DMs. This was done by choosing the highest-S/N burst in each group containing more than 4 candidates with $\mathrm{S/N}>5$, and rejecting the rest. The remaining candidate bursts were assessed using the FETCH\footnote{\url{https://github.com/devanshkv/fetch}} \citep{agarwal_2020_mnras} deep-learning-based classifier. FETCH is trained to classify candidates based on their frequency-time image (i.e., dynamic spectrum) and their DM-time image (i.e., bow-tie plot). Since the S/N drops steeply with changing DM at the low radio frequencies of our observations \citep{cordes_2003_apj}, the characteristic DM-time bow-tie variation would not be visible at large DM deviations. Hence, we modify the DM range for the DM-time image to go from $[0.9\text{DM}, 1.1\text{DM}]$ instead of the default $[0, 2\text{DM}]$ used in FETCH. All FETCH candidate bursts with a \texttt{single\_pulse\_search.py} reported $\mathrm{S/N}>6$ --- and with a mean probability of being astrophysical $>50$\% from the 11 available FETCH deep-learning models --- were visually inspected to remove possible contamination by residual RFI. 

In our observations, we detect 11 new bursts, bringing the total number of bursts detected from \rthree\ by LOFAR to 29. The bursts have been labeled B1 through B29 (of which B1$-$B18 were reported in \citetalias{pleunisetal}), in chronological order of barycentric arrival time. The bandpass-corrected dynamic spectra and time series profiles of bursts B19$-$B29 are plotted in Figure~\ref{fig:burstfamily} and the corresponding burst parameters can be found in (Table~\ref{tab:B19_to_B29_burst_properties}).

\begin{table*}
  \centering
  \footnotesize
  \caption{Burst parameters. See Section~\ref{sec:results} for a description of how parameters were determined. For all bursts, arrival times and burst widths are computed for $\mathrm{DM}=348.772$\,\pcpercc\ \citep{nimmo_2021_natas}. 
  \label{tab:B19_to_B29_burst_properties}
  }
  \begin{tabular}{lccccccccc}
    \hline
    Burst & \multicolumn{2}{c}{Barycentric arrival time (TDB) at $\nu=\infty$} & $\phi$ & Gaussian burst width$\mathrm{^{a}}$ & $\nu_\mathrm{low}$$\mathrm{^{b}}$ & $\nu_\mathrm{high}$$\mathrm{^{b}}$ & S/N$\mathrm{^{c}}$ & Fluence$\mathrm{^{d}}$  \\
    ID & (UTC) & (MJD) & & (ms) & (MHz) & (MHz) & & (Jy ms) \\
    \hline

B19 & 2021-03-05T17:05:36.672 &     59278.71222943 & 0.68   &   $46\pm{1}$    & 145.2 & 188.0 &  29.3 &       $398\pm{44}$ \\
B20 & 2021-03-07T10:59:23.424 &     59280.45790944 & 0.79   &   $38.4\pm{0.9}$    & 120.1 & 155.8 &  25.0 &       $412\pm{50}$ \\
B21 & 2021-06-09T06:08:41.856 &     59374.25604087 & 0.53   &   $97\pm{1}$    & 109.9 & 131.1 &  13.1 &       $438\pm{59}$ \\
B22 & 2021-06-09T06:20:04.416 &     59374.26394068 & 0.53   &   $66\pm{4}$    & 113.0 & 134.6 &  11.6 &       $346\pm{54}$ \\
B23 & 2021-06-09T06:22:10.560 &     59374.26540131 & 0.53   &   $33\pm{1}$     & 132.6 & 188.0 &   6.4 &       $162\pm{32}$ \\
B24 & 2021-10-20T00:04:45.984 &     59507.00331102 & 0.66   &   $47\pm{1}$    & 113.0 & 154.2 &  31.3 &       $402\pm{80}$ \\
B25 & 2021-11-20T20:55:33.888 &     59538.87192058 & 0.61   &   $15\pm{6}$ and $26\pm{3}$   & 152.7 & 188.0 &  16.0 &  $343\pm{69}$ \\
B26 & 2022-08-08T06:50:19.680 &     59799.28495172 & 0.56   &    $5\pm{9}$    & 158.9 & 177.8 &   3.1 &       $39\pm{12}$ \\
B27 & 2022-08-11T10:33:17.856 &     59802.43978837 & 0.76   &   $19\pm{7}$    & 136.2 & 181.7 &   6.5 &       $176\pm{35}$ \\
B28 & 2022-09-28T00:28:48.000 &     59850.01999957 & 0.67   &   $36.5\pm{0.6}$    & 139.6 & 188.0 &   8.6 &   $88\pm{18}$ \\
B29 & 2022-09-28T00:45:41.472 &     59850.03172779 & 0.67   &   $24\pm{9}$    & 135.4 & 188.0 &   5.1 &       $141\pm{44}$ \\
  \hline
  \multicolumn{8}{l}{$\mathrm{^{a}}$ Burst width (FWHM of Gaussian), for a Gaussian function with exponential tail fitted to the time series.}\\
  \multicolumn{8}{l}{\hspace{0.6em} See Section~\ref{ssec:scatteringdrift} for fitting method used.}\\
  \multicolumn{8}{l}{$\mathrm{^{b}}$ $\nu_\mathrm{low}$ and $\nu_\mathrm{high}$ are determined by fitting a Gaussian profile to the spectrum of the burst and taking}\\
  \multicolumn{8}{l}{\hspace{0.6em} the $\pm 3\sigma$ limits of the fitted profile. If the $\nu_\mathrm{low}$ or $\nu_\mathrm{high}$ thus obtained are beyond the edges}\\
  \multicolumn{8}{l}{\hspace{0.6em} of the observing band, the edge of the observed band occupied by the burst is quoted.}\\
  \multicolumn{8}{l}{$\mathrm{^{c}}$ Profile integrated S/N of the time series profile, after summing over the frequency range occupied by the burst.}\\
  \multicolumn{8}{l}{\hspace{0.6em} Note that this value is different from the peak S/N and our detection metric \texttt{single\_pulse\_search.py}-based S/N.}\\
  \multicolumn{8}{l}{$\mathrm{^{d}}$ Fluence determined using only the frequency channels occupied by the burst.} \\
  \multicolumn{8}{l}{\hspace{0.6em} The procedure followed for fluence calculation and fluence error estimation is the same as described in \citetalias{pleunisetal}.}

  \end{tabular}
\end{table*}


\section{Analysis and Results}\label{sec:results}

\begin{table*}
  \centering
  \footnotesize
  \caption{Measurements of propagation effects for the new bursts presented here. See also Table~\ref{tab:B1_to_B18_burst_properties} for a re-fitting of the bursts previously published in \citetalias{pleunisetal}.} 
  \begin{tabular}{lcccccc}
    \hline
    Burst & {$\tau_s$ (at 150\,MHz)}$\mathrm{^{a}}$ & Drift rate (ACF method)$\mathrm{^{b}}$ &  Drift rate (time of arrival method)$\mathrm{^{c}}$ & RM$\mathrm{^{d}}$ & RM$_\mathrm{iono}$ & L/I  \\
    ID & (ms) & (\mhzperms)  & (\mhzperms) & (\radpermsq) & (\radpermsq) & (\%) \\
    \hline
B19  &  71$_{ -5}^{+6}$ &  $-0.5\pm{0.2}$& $-1.6\pm{0.3}$ &$-115.5\pm{0.3}$  &   $+0.62\pm{0.04}$ &   $39\pm{6}$   \\
B20  &  58$_{ -5}^{+2}$ &  $-0.4\pm{2.7}$  &  $-1.6\pm{1.1}$ &$-116.6\pm0.3$   &   $+0.72\pm{0.06}$  &   $41\pm{9}$  \\
B21  &  32$_{ -3}^{+2}$ &  $-0.03\pm{0.06}$ & -- & --  &   --    &   $<18\pm{26}$ \\
B22  &  69$_{-11}^{+98}$ &  $-0.02\pm{0.2}$& --  &  --  &   --  &   $<24\pm{18}$ \\
B23  &  -- &  --    &   --  &  $-105.8\pm{0.3}$   &   $+0.23\pm{4}$    &   $34\pm{16}$    \\
B24  &  44$_{-5}^{+2}$ &  $-0.3\pm{0.2}$ & $-1.4\pm{0.7}$ & --   &   --     &   $<16\pm{11}$    \\
B25  &   --   &  $-0.2\pm{1}$ & --   &  --   &   --   &   $<12\pm{24}$    \\
B26  &  --   &  --   & -- &  --   &   --   &   --    \\
B27  &    --     &  --   &  --&  $-54.4\pm{0.4}$ &   $+1.24\pm{5}$    &   $50\pm{15}$       \\
B28  &   -- &  --    & --   &  --   &   --    &   $<26\pm{29}$    \\
B29  &    --   &  --    & --   &  --   &   --    &   $<27\pm{27}$    \\
   \hline
   \multicolumn{5}{l}{$\mathrm{^{a}}$ Scattering timescales are only quoted for single-component bursts with $\mathrm{S/N}>10$.} \\ 
   \multicolumn{5}{l}{\hspace{0.6em} The errors quoted here are only statistical from the MCMC fit.} \\
   \multicolumn{5}{l}{\hspace{0.6em} We incorporate deviations in scattering from MCMC fits to simulated bursts}\\
   \multicolumn{5}{l}{\hspace{0.6em} to account for covariance with drift rates as an additional source of error in Figure~\ref{fig:scattering}.}\\
   \multicolumn{5}{l}{\hspace{0.6em} See Section~\ref{ssec:scatteringdrift} for a full description.}\\
   \multicolumn{5}{l}{$\mathrm{^{b}}$ Drift rates using the ACF method are only quoted for bursts with $\mathrm{S/N}>10$.}\\
   \multicolumn{5}{l}{$\mathrm{^{c}}$ Drift rates using the time of arrival method are only quoted for bursts with $\mathrm{S/N}>20$.} \\
    \multicolumn{5}{l}{\hspace{0.6em} The S/N limit here is higher than for the ACF method since we are more limited by S/N by dividing} \\
    \multicolumn{5}{l}{\hspace{0.6em} individual bursts into subbands to calculate the times of arrival at each subband.}\\
   \multicolumn{5}{l}{$\mathrm{^{d}}$ Observed RM, uncorrected for ionospheric contribution. Error in RM calculated as $\frac{\mathrm{FWHM}_{\mathrm{FDF}}}{2*\mathrm{S/N}}$ .}\\
   \end{tabular}
   \label{tab:prop_effects}
\end{table*}
\subsection{Bursting activity}\label{ssec:activity}

Figure~\ref{fig:activity} shows the barycentric times of arrival, total bursts per 16.33-day activity window, and burst rates of all 29 LOFAR bursts along with the CHIME/FRB bursts detected since the first LOFAR observations of \rthree\ began (2019 June 6), plotted against the activity phase. Activity phases are determined by setting MJD~58369.40 at phase $\phi_0$ (i.e, $\phi$ = 0.0), such that $\phi = 0.5$ is the mean of the folded phases of the CHIME/FRB bursts in \citetalias{pleunisetal}. We find that the 11 new bursts fall within the previously observed $4.1$-day ($25$\% of the 16.33-day cycle) LOFAR activity window (\citetalias{pleunisetal}) between activity phases $0.53<\phi<0.79$. 

In order to examine if the properties of the activity window remain stable, accounting for the varying exposure of the LOFAR observations, we repeat the Markov-Chain Monte Carlo (MCMC) analysis from \citetalias{pleunisetal} (their Section~3.5, Figure~10) for bursts from different time ranges. We briefly repeat the description of the analysis here, which uses \texttt{emcee} by \citet{foremanmackey_2013_pasp} for the MCMC analysis. For each pair of the central window phase $\phi_{\mathrm{c}}^{\mathrm{LOFAR}}$ and activity window width $w$, $N$ burst arrival times are randomly drawn in the windows of allowed activity in the specified epoch, assuming a uniform burst rate across activity cycles and across the allowed activity phases. The number of bursts $N$ for a given epoch is based on the known average burst rate in the window of observed activity. This analysis assumes that there is no significant derivative of the activity period, and random values for $\phi_{\mathrm{c}}^{\mathrm{LOFAR}}$ and $w$ are drawn from uniform distributions with $ 0.0 \leq \phi_{\mathrm{c}}^{\mathrm{LOFAR}} \leq 1.0$ and $4 \leq w \leq 10$\,day. We thus arrive at a probability distribution function of obtaining bursts that fall within the observed phases, given the number of realizations out of 10,000 draws in which all drawn bursts fall within the observed activity phases. The posterior distributions were determined by using 20,000 steps of 32 walkers until the chains converged. The best fit parameters were obtained from the posterior distributions after discarding a burn-in phase of 1,000 steps and thinning the chains by 5 steps.

\begin{figure*}
  \centering

  \includegraphics[width=0.44\textwidth]{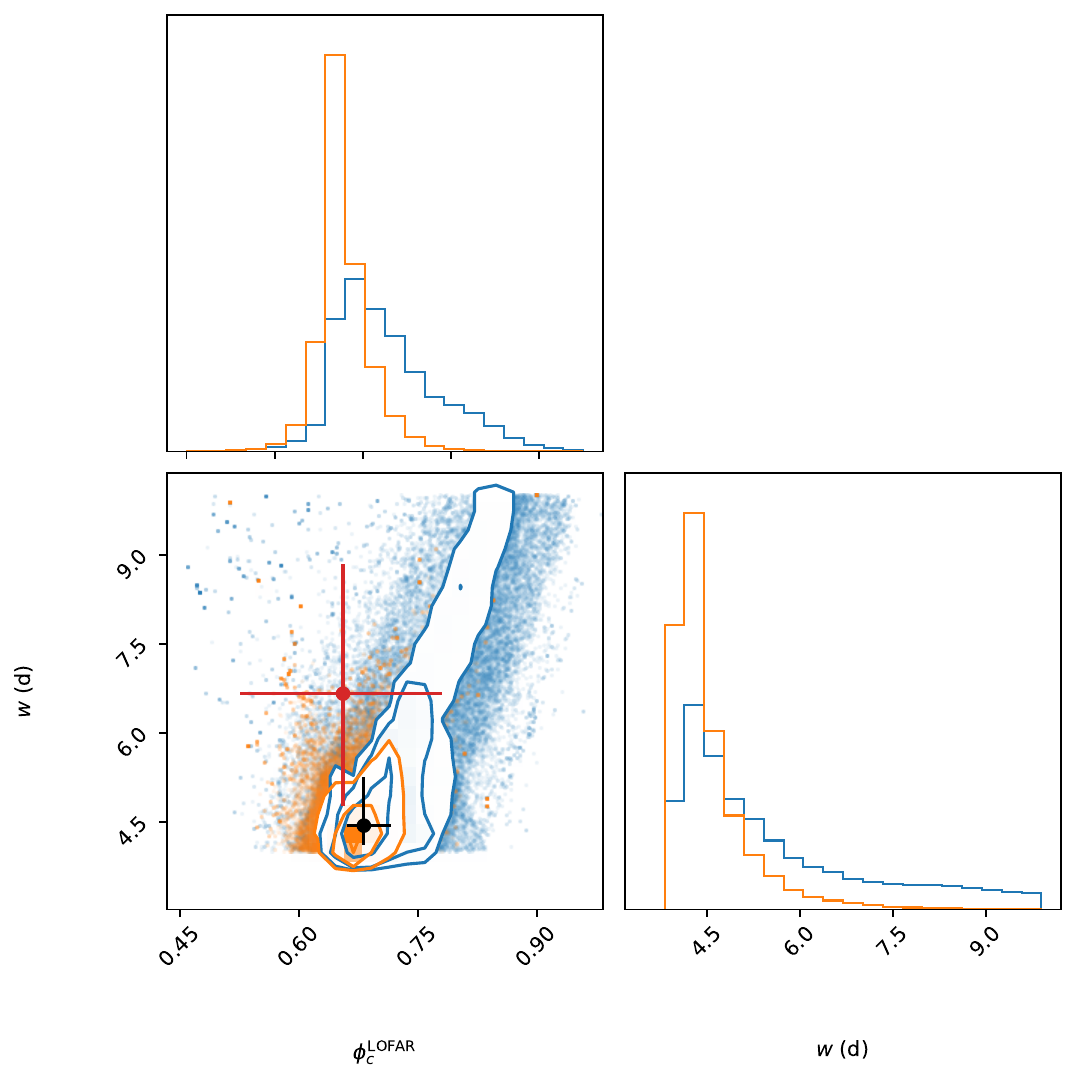}
  \caption{Corner plot from the MCMC analysis showing constraints on the width of the LOFAR activity window $w$, and the activity phase at which it is centered $\phi_{\mathrm{c}}^{\mathrm{LOFAR}}$, for different time ranges. These constraints are derived from accounting for the telescope exposure across all activity phases in the 16.33-day period. With $\sim2.5$ times the total exposure time since \citetalias{pleunisetal}, and  49.8\% of this spread outside the observed activity window to date, we obtain stronger constraints on the width of the window. \textit{In blue}: Constraints on $w$ and $\phi_{\mathrm{c}}^{\mathrm{LOFAR}}$ using only the first 18 bursts, and exposures up until MJD~59088, as published in \citetalias{pleunisetal} (their Figure~10). \textit{In orange}: Constraints on $w$ and $\phi_{\mathrm{c}}^{\mathrm{LOFAR}}$ using all bursts and observations up to MJD~59854 (top-left sub-figure). \textit{In black}: Constraints obtained for exposures before MJD~59355, at which the first burst with a $\sim10$\% change in RM is observed. \textit{In red}: Constraints obtained for exposures after (and including) MJD~59355, at which the first burst with a $\sim10$\% fractional change in RM is observed. This shows that properties of the activity window at LOFAR have not shifted since the observed secular trend in RM began. }
  \label{fig:window_simulations}
\end{figure*}

These simulations to fit for $\phi_{\mathrm{c}}^{\mathrm{LOFAR}}$ and $w$ in the $110-188$\,MHz band are done for the following four time ranges and the distributions with fits shown in Figure~\ref{fig:window_simulations}: (a) From MJD~58640 to MJD~59087, with 18 detected bursts. The best fit parameters are  $\phi_{\mathrm{c}}^{\mathrm{LOFAR}} = 0.72^{+0.07}_{-0.04}$ and $w = 5.02^{+2.34}_{-0.78}$ (restating the window properties from \citetalias{pleunisetal}); (b) From MJD~58640 to MJD~59854, with 29 detected bursts (this includes all LOFAR observations of and bursts from \rthree~to date). We find that the most probable activity window width is $w = 4.32^{+0.67}_{-0.24}$\,day, which is more tightly constrained than previously found in \citetalias{pleunisetal}. The window is centered at a phase of $\phi_{\mathrm{c}}^{\mathrm{LOFAR}} = 0.67^{+0.03}_{0.02}$, which is earlier than the previously obtained value by nearly a day (although still consistent within errors). This is still significantly offset from the CHIME/FRB window centered at a phase $0.5^{+0.058}_{-0.058}$ \citep{chime_2020_natur_periodicactivity}, lending support to the chromatic modulation of activity; (c) From MJD~58640.4 to MJD~59355, during the stochastic RM variations of $2-3$ \radpermsq. The best fit window parameters are $\phi_{\mathrm{c}}^{\mathrm{LOFAR}} = 0.68^{+0.03}_{-0.02}$ and $w = 4.44^{+0.83}_{-0.33}$; (d) From MJD~59355 to MJD~59854, during the observed secular trend in RM in the $400-800$-MHz CHIME/FRB band \citep[][see Section~\ref{ssec:pol}]{Mckinven_2023_ApJ_R3}. The best-fit window parameters are $\phi_{\mathrm{c}}^{\mathrm{LOFAR}} = 0.66^{+0.13}_{-0.13}$ and $w = 6.66^{+2.19}_{-1.90}$. The implications of the activity window properties remaining consistent before and after the start of the secular RM trend, i.e., cases (c) and (d), are discussed later in Section~\ref{sec:disc}. 

 Burst rates are calculated across time --- for each 16.33-day activity period (only considering exposures within the observed LOFAR activity window), and across activity phase --- including all cycles during which we had observations. Assuming that the bursts detected within the activity window follow a Poisson process, we calculate the $1\sigma$ errors following \citet{Gehrels_1986_ApJ}. The obtained rates are plotted in panels 1c and 2c of Figure~\ref{fig:activity}, with upper limits for epochs with no detected bursts. We find that the burst rates are largely consistent between activity cycles within $1\sigma$ (panel~1c, Figure~\ref{fig:activity}). By distributing our observations across all activity phases (as described in Section~\ref{sec:obs}), we are able to place upper limits on the LOFAR burst rate throughout the activity cycle, outside of the expected LOFAR activity window. The average burst rate and the 3-$\sigma$ Poissonian errors during the activity window we obtain in this work, i.e., case~(b), is $0.17^{+0.28}_{-0.09}$\,h$^{-1}$, which differs from the upper limit on burst rate outside the expected window of $0.006$\,h$^{-1}$, by over 3$\sigma$ (see also panel~2c of Figure~\ref{fig:activity}).   

\subsection{Scattering timescale and sub-burst drift rate variations}\label{ssec:scatteringdrift}

Scattering timescale fits can be covariant with other frequency-time properties of the bursts such as DM, sub-burst drift rate \citep{hessels_2019_apjl}, and the intrinsic width of the underlying burst (as it was emitted). \citet{Sand_2022_ApJ} constrain DM variations to $< 1$\,\pcpercc\ at 800\,MHz for bursts in January 2020 using structure maximization, while \citet{Mckinven_2023_ApJ_R3} find $< 0.8$\,\pcpercc\ variations during a 9-month period in 2021 (that overlaps with our observing period). Using a different approach to find structure optimized DM, \citet{Lin_2022_arXiv} claim DM variations of $\sim$1.5\,\pcpercc. However, the optimized DM using their \texttt{DM-power} algorithm on the highest S/N burst in their sample agrees with the microsecond, structure-optimized DM from \citet{nimmo_2021_natas} to within $0.005$\,\pcpercc. There is a nine-month difference between these two consistent DM measurements. \citet{pastormarazuela_2021_natur} find a structure optimizing DM $= 348.75 \pm 0.12$\,\pcpercc~for APERTIF bursts at L-band, and find that this limits the rate of change of DM to less than 0.05\,\pcpercc~per year, with bursts separated by up to $\sim17$ months from the \citet{nimmo_2021_natas} measurement.

We assume here that bursts from this source are intrinsically made up of multiple sub-bursts that drift towards lower frequencies as time progresses. In the limit of low S/N, the individual sub-bursts blend together to produce one `blob' of emission \citep[see, e.g., Figure~7 in ][]{gourdji_2019_apjl}. When measuring DMs in this low S/N limit, one is biased high due to any dedispersion algorithm (optimizing for burst S/N or burst structure) preferring to superimpose the sub-bursts that are drifting downwards in frequency. This is seen in the middle panel of Figure~3 of \citet{chime_2020_natur_periodicactivity} and in Figure~6 of \citet{Lin_2022_arXiv}, where the per-burst best-fit DMs are shown for \rthree. Notably, the best-fit DMs derived from high-time-resolution data in \citet{chime_2020_natur_periodicactivity} are all within 0.1\,\pcpercc~of the \citet{nimmo_2021_natas} DM value.  

Given the absence of evidence for significant DM variations, we choose to dedisperse all bursts to the same structure-optimized DM $= 348.772$\,\pcpercc\ from microsecond burst structure measured by \citet{nimmo_2021_natas}. Using a DM value optimized for microsecond structure circumvents the issue of covariance of the DM with spectro-temporal changes of the burst profile, particularly exacerbated by the obscuring of structure by scattering at lower frequencies. Upon dedispersing the dynamic spectra, we can measure the times-of-arrivals of the burst at different parts of the burst bandwidth, to check for residual quadratic delays from dispersion. Figure~\ref{fig:toa_vs_freq} shows these delays versus frequency for the brightest bursts (S/N $> 20$). Since the times-of-arrivals progressively move to later times at lower frequencies for all the bursts, any variations in DM must all be in the same positive direction compared to DM $= 348.772$\,\pcpercc, which we find to be unlikely. In the case of stochastic DM variations, we would expect that the DM randomly fluctuates around the used DM, rather than in one direction preferentially. We see that deviations in DM can be no greater than $+0.25$\,\pcpercc, although we note that most of the delays in individual bursts appear non-quadratic and thus interpret it as sub-burst drifting instead. The delays being larger for bursts at the bottom of the band can be explained by the expected linear scaling of drift rate with frequency, as we discuss in more detail below.

We are unable to perform a joint-fit for the scattering timescales and drift rates due to insufficient S/N of the bursts in the frequency-time dynamic spectra. Hence, we fit for these two quantities separately and calculate the uncertainties due to drift-scattering covariance using simulated bursts. 

Downward drifting sub-bursts towards later times and lower frequencies have been seen to be characteristic of repeating FRBs \citep{hessels_2019_apjl,fonseca_2020_apjl}. The two distinct components visible in the dynamic spectrum of B25 show that this `sad trombone effect' is manifest in bursts from \rthree\ even at LOFAR frequencies. In the case of bursts that do not show distinct sub-bursts, yet show delays towards lower frequencies in the dynamic spectra post dedispersion, we assume that the downward drifting sub-structures have been obscured by low S/N, or shorter waiting times, or large scattering timescale, or any combination of the above.
To measure the linear drift rate of the bursts ($\frac{d\nu}{dt}$ or $\dot{\nu}$), we fit a two-dimensional Gaussian ellipsoid to the two-dimensional auto-correlation function (ACF) of the dedispersed dynamic spectra of the bursts (see Figure~\ref{fig:drifts_familyplot} for the 2D ACF fits obtained). The 2D-Gaussian ellipsoid is parameterized as follows, after applying a counterclockwise rotational transformation by an angle $\theta$:
\begin{align}
    G\left(x,y\right) & = A\exp\left\{-\frac{1}{2}\left[x^2\left(\frac{\cos^{2}\theta}{\sigma_{x}^2}+\frac{\sin^{2}\theta}{\sigma_{y}^2}\right)\right.\right.\nonumber\\
    & \left.\rule{0mm}{0mm}\left.+2xy\sin\theta\cos\theta\left(\frac{1}{\sigma_{x}^2}-\frac{1}{\sigma_{y}^2}\right)+y^2\left(\frac{\sin^{2}\theta}{\sigma_{x}^2}+\frac{\cos^{2}\theta}{\sigma_{y}^2}\right)\right]\right\}
    \label{eq:Gaussian_SM}
\end{align}

Here, $\sigma_{x}$ and $\sigma_{y}$ are the standard deviations of the Gaussian ellipsoid along orthogonal axes after the rotational transform is performed. The inclination of this ellipsoid $\mathrm{cot}(\theta)$, gives us the drift rate in \mhzperms. We use the \texttt{lmfit} Python package to perform a non-linear least-squares fit for the free parameters A, $\sigma_{x}$, $\sigma_{y}$ and $\theta$. The uncertainty on $\theta$ is calculated by taking the square-root of the variance from the fit, summed with the square-root of the reduced-$\chi^{2}$ of the residuals from the fit as an additional source of error. Previously, the magnitude of the drift rate has been seen to roughly linearly decrease with decreasing frequency for \rthree\ as well as \rone\ \citep{Chamma_2021_MNRAS,hessels_2019_apjl}. For \rthree, we are able to recover this relation of decreasing drift rate magnitudes with frequency, even within the 78-MHz LOFAR bandwidth (Figure~\ref{fig:driftrate}). We obtain the relation $\dot{\nu}=(-0.02 \pm 0.01) \nu_{c}$ from a least-squares fit of a line to the mean value of the drift rates within the LOFAR band (without weighting the drift rates by their errors). Considering measurements from higher frequencies \citep{chawla_2020_apjl, Chamma_2021_MNRAS, pastormarazuela_2021_natur, Sand_2022_ApJ}, the drift rate scales as $\dot{\nu}=(-0.027 \pm 0.005) \nu_{c}$ for this source between 110\,MHz and 1520\,MHz, which agrees with the $\dot{\nu}=-0.02 \nu_{c}$ relation obtained by \citet{Sand_2022_ApJ} for drift rates measured between $400-1500$\,MHz. 

As a second method to measure the drift rate, we use the slope of a linear fit to the times-of-arrival at different frequencies (see Figure~\ref{fig:toa_vs_freq}). The drift rates measured this way are also plotted in Figure~\ref{fig:driftrate}, and quoted in Tables~\ref{tab:prop_effects} and \ref{tab:B1_to_B18_burst_properties}. They are found to be consistent within errors for most bursts where the drift rates were measured using the two-dimensional ACFs, and they follow an inverse linear trend with frequency with a slope of $-0.04\pm0.03$.

Inhomogeneities in the medium through which the bursts propagate results in a geometric time delay between rays from the burst received at the telescope, manifesting as a temporal broadening of the pulse. This pulse broadening due to multi-path scattering can be described by a one-sided exponential function in the case of an infinitely extended, thin screen. The resulting time ($t$) variant burst profile $\mathrm{I}(t)$ is a convolution of the pulse broadening function with the underlying emitted burst profile:
\begin{align} \label{eq:scattering}
    \mathrm{I}(t) = \exp(t/\tau_\mathrm{s}) * \Gamma(t, \sigma_{t}) 
\end{align} 

Here, $\tau_\mathrm{s}$ is the scattering timescale and $\Gamma(t, \sigma_{t})$ is the underlying burst profile before scattering, which we assume to be a Gaussian with standard deviation $\sigma_{t}$. Scattering timescale varies with frequency as $\tau_\mathrm{s} \propto \nu^{\alpha}$, where generally scattering index $\alpha$ is between $-4$ \citep[for a thin-screen scattering model with Gaussian inhomogeneities]{Williamson_1974_MNRAS, Lang_1971_ApJ} and $-4.4$ \citep[Kolmogorov spectrum]{Lee_1975_ApJ}. However, it is possible that scattering scales with frequency with a different $\alpha$. To test this we measure scattering timescales at different subbands of individual bursts by performing a \texttt{scipy} least-squares fit of a model of the burst as given in Equation~\ref{eq:scattering} to the different subbands (Figure~\ref{fig:scattering_index_fit}), and measure varying indices between $-4.5 \leq \alpha \leq -1.7$. Given the large errors on the indices (see Figure~\ref{fig:scattering_index_fit}), we choose to scale the scattering timescales measured below to an $\alpha = -4.0$ to reference them to 150\,MHz, and include the deviation in scattering timescale if scaled using $\alpha \in [-4.4, -3.6]$ as a part of the error. The $\alpha$ fit for burst B6 has the smallest error associated and agrees with $\alpha = -4.0$.

\begin{figure*}
  \centering
  \includegraphics[width=0.9\textwidth]{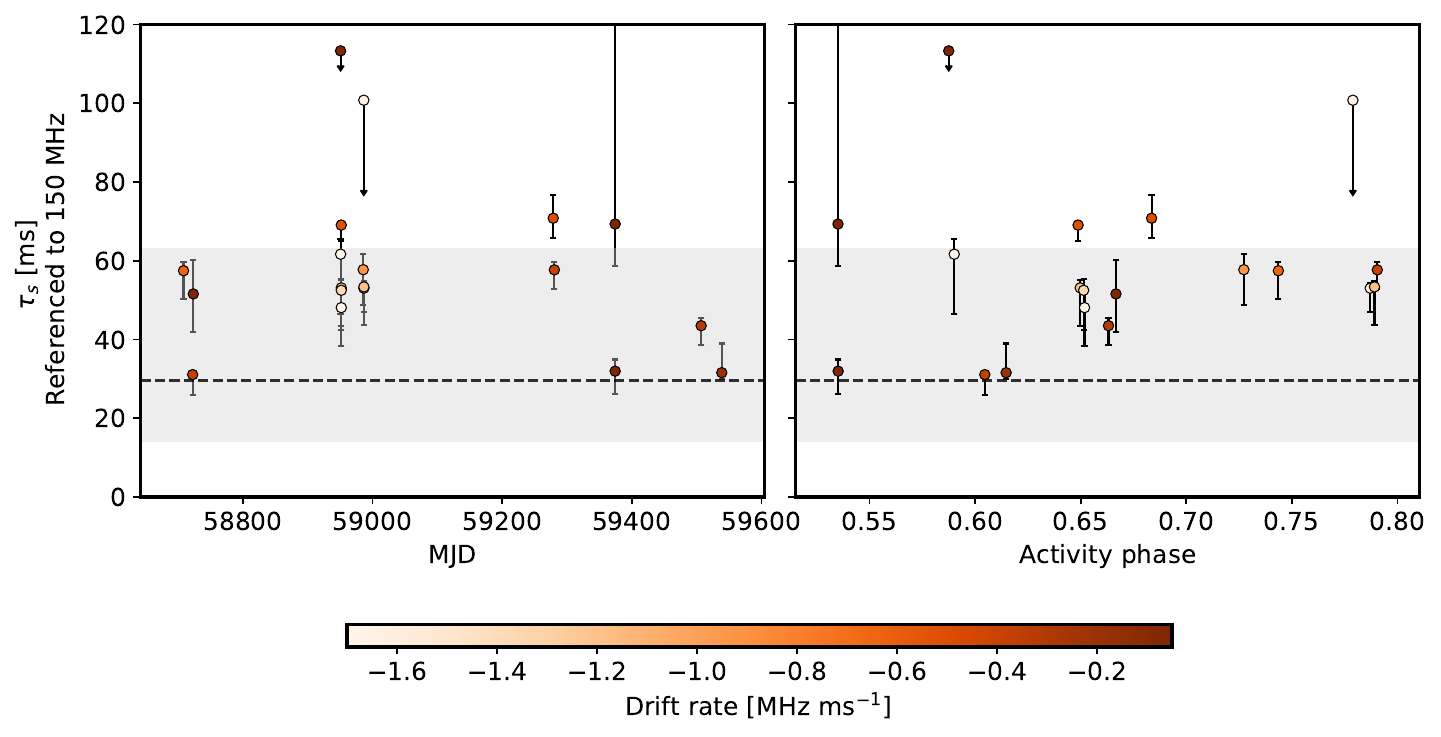}
  \caption{Variation of scattering timescales of the bursts with time and activity phase, obtained by fitting a Gaussian convolved with an exponential tail to the time profiles of the bursts. Only values for bursts with a $\mathrm{S/N}>10$ are marked in the plot. All scattering times have been referenced to $150$\,MHz, assuming $\tau_{s} \propto \nu^{-4}$. The error bars are a quadrature sum of the  1-$\sigma$ statistical errors from the MCMC fitting, the errors derived from Figure~\ref{fig:driftscatteringfit} for the covariance between scattering and drift rate (as described in Section \ref{ssec:scatteringdrift}), and difference in scattering timescale if the values were scaled using $\tau_{s} \propto \nu^{[-3.6\leq \alpha \leq -4.4]}$ when referenced to 150\,MHz. The data points are coloured by their corresponding fitted drift rates. We include measurements for all detected \rthree\ LOFAR bursts B1$-$B29, i.e., including bursts reported in \citetalias{pleunisetal}. The black dashed line represents the expected line-of-sight scattering from the Milky Way disk based on the NE2001 model, after accounting for plane-wave scattering of extragalactic bursts, and scaling the expected value at 1\,GHz using $\tau_{s} \propto \nu^{-4}$. The gray shaded region shows the Milky Way ISM expected scattering contribution (NE2001 model) for frequency scaling indices in the range $[-3.6 \leq \alpha \leq -4.4]$.}
  \label{fig:scattering}
\end{figure*}

\begin{figure*}
  \centering
  \includegraphics[width=0.75\textwidth]{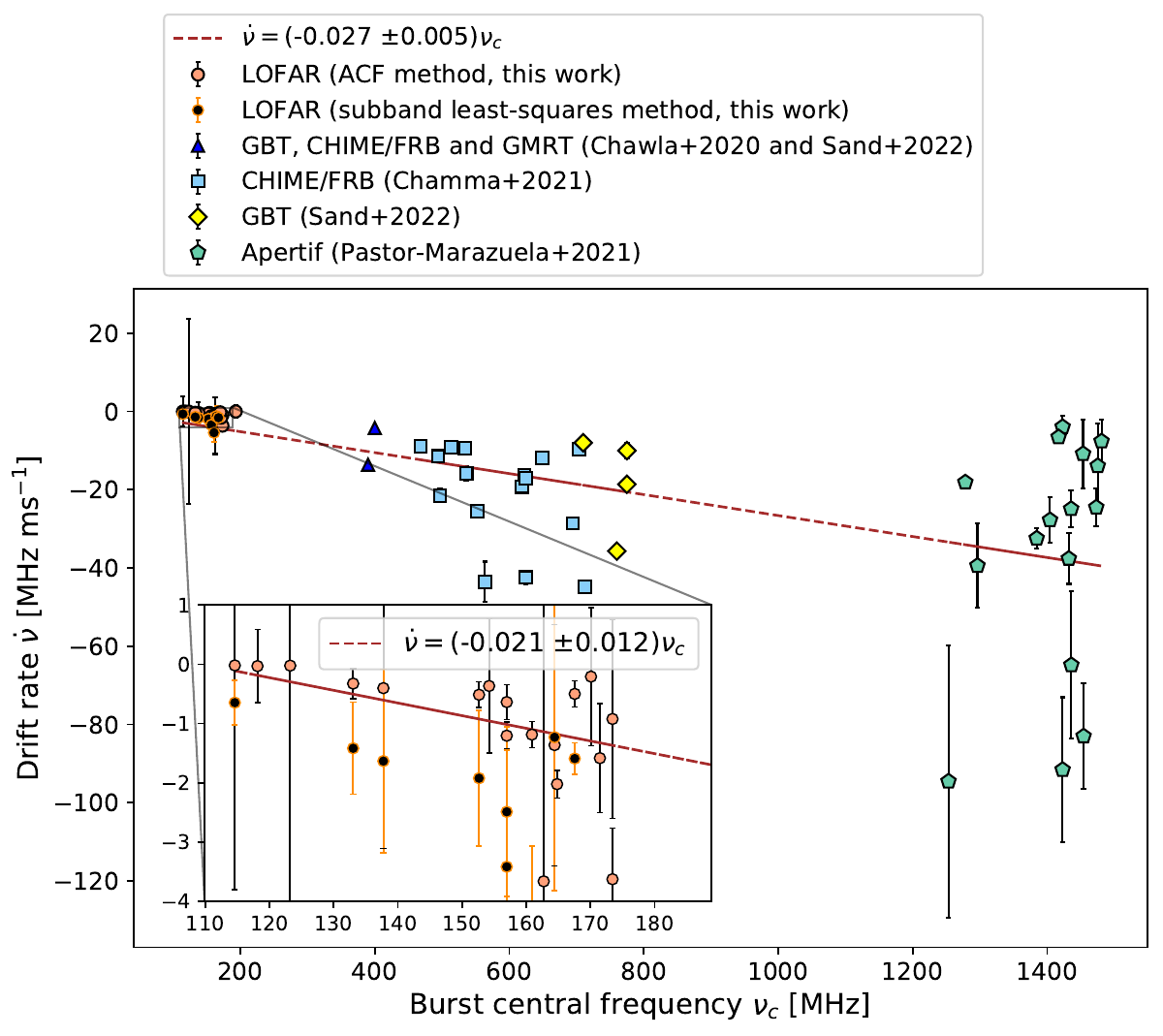}
   \caption{Drift rates versus the centre frequency of the 
  bursts in the LOFAR band (inset: orange circle markers (ACF method), black circle markers (time of arrival method)), compared to the measured rates at higher frequencies --- GBT, CHIME/FRB, and GMRT $300-500$\,MHz \citep{chawla_2020_apjl, Sand_2022_ApJ}: dark blue triangle markers; CHIME/FRB $400-800$\,MHz \citep{Chamma_2021_MNRAS}: blue square markers; GBT $600-1000$\,MHz \citep{Sand_2022_ApJ}: yellow diamonds; Apertif $1220-1520$\,MHz \citep{pastormarazuela_2021_natur}: green pentagonal markers. The 1-$\sigma$ errors of the drift rates are included. We measure drift rates for all detected \rthree\ LOFAR bursts B1$-$B29, i.e., including bursts reported in \citetalias{pleunisetal}.
  We obtain the relation $\dot{\nu}=-0.02 \pm 0.01 \nu_{c}$ from a least-squares fit of a line to the drift rates in the LOFAR band. 
  The linear decrease in drift rate (using the ACF method) with the observed frequency 
  measured within the LOFAR band is consistent within errors with the best-fit linear relation 
  of $\dot{\nu}=(-0.027 \pm 0.005) \nu_{c}$ for this source between 110\,MHz and 1520\,MHz. Neither of these fits weight the drift rates by their individual uncertainties.}
  \label{fig:driftrate}
\end{figure*}

Scattering timescales were measured by fitting the total intensity profiles of each burst with a symmetric Gaussian envelope ( $\Gamma(t, \sigma_{t})$) convolved with an exponential tail first by a least-squares fit using the \texttt{scipy} Python package. The obtained values were used as the initial values for a least-squares fit using the MCMC method (with the \texttt{emcee} Python package; \citealt{foremanmackey_2013_pasp}). The results of the fitting algorithm can be compared with the burst profile in Figure~\ref{fig:scattering_residuals}, which also shows the residuals after subtracting the model fit from the data. One Gaussian component was assumed for all bursts except B25 where the downward-drifting substructures are distinguishable in the dynamic spectrum (Figure~\ref{fig:burstfamily}). We assume that scattering time scales with frequency as $\nu^{-4}$, and only measure it for bursts with $\mathrm{S/N}>10$. Since our scattering fits to bursts B7 and B14 (see Figure~\ref{fig:scattering_residuals}) are affected by the presence of possible secondary peaks in their frequency-integrated time profiles, we consider the measured value for these bursts to only be an upper limit. The measured scattering timescales (Tables~\ref{tab:prop_effects} and \ref{tab:B1_to_B18_burst_properties}), referenced to $150$\,MHz, span a range between $\sim30$\,ms and $\sim70$\,ms (excluding B7 and B14), as seen in Figure~\ref{fig:scattering}.

To examine the effect of drifting on the scattering timescales obtained through such a fitting process, we fit scattering timescales to the time profiles of simulated bursts with non-zero drift rates in their dynamic spectrum, using the same fitting process. The dynamic spectrum of each of the simulated bursts was modeled as a single tilted 2D Gaussian, as described in Equation~\ref{eq:Gaussian_SM} (since most of the detected bursts do not show distinct components), after which the channels containing the burst had their Gaussian profiles convolved with an exponential scattering tail as described in Equation~\ref{eq:scattering}. The simulated bursts had the following input properties:
\begin{itemize}
    \item Standard deviation of the Gaussian intrinsic burst width in time $\sigma_{t} = 16$\,ms, the median value from the instrinsic widths of B1$-$B29.   
    \item Standard deviation of the Gaussian profile extent of the burst in frequency $\sigma_{\nu} = 6$\,MHz, the median value from the frequency extents of B1$-$B29. 
    \item Scattering timescales $\tau_{s}$ referenced to 150\,MHz, sampled uniformly in the range [0\,ms, 150\,ms], at the central frequency of the burst.
    \item Drift rates $\dot{\nu}$ sampled uniformly in the range [$-1.7$\,\mhzperms,$-0.01$\,\mhzperms]
    \item Central frequency of the burst dictated by the input drift rate, as $\nu_{c}= (\dot{\nu}-2.336)/0.021$, the best-fit relation from Figure~\ref{fig:driftrate}. 
    \item Time series S/N (signal summed over the extent of the burst and divided by the square root of the width of the burst in time) of 20.
    \item Equal fluences, by normalizing the time series by the area under the burst.
\end{itemize} 

Figure~\ref{fig:driftscatteringfit} shows the fractional difference between the fitted scattering timescale ($\tau_{s(\mathrm{fit})}$) and simulated scattering timescale ($\tau_{s(\mathrm{input})}$), for varying input drift rates. We treat this fractional difference as the covariance of the fitted scattering timescales with drift rate. The measured drift rates for bursts B1$-$B29 are then matched with the input drift rates of the simulated bursts, and the measured scattering timescales are matched with the fitted scattering timescales from the simulation. An additional source of one-sided-error (based on whether the fit is over/under-estimated in relation to the input scattering in the simulated bursts) on the measured scattering timescales ($\tau_{s(\mathrm{meas})}$) of bursts B1$-$B29, given the measured drift rate is derived as : 
\begin{equation*}
\frac{\tau_{s(\mathrm{fit})}-\tau_{s(\mathrm{input})}}{\tau_{s(\mathrm{fit})}} \cdot \tau_{s(\mathrm{meas})}
\end{equation*}

\subsection{Fractional polarization and RM variations}\label{ssec:pol}

LOFAR has dual-linear polarized feeds that are stationary on the ground and aligned along the South-East to North-West and South-West to North-East directions. Depending on the elevation of the source at the time of observation, the electromagnetic waves from the burst will be projected on to the stationary telescope feeds at an angle, degradation of signal intensity and distortion of the polarization signal \citep{Noutsos_2015_A&A}. We apply a Jones calibration matrix calculated from a beam-model of LOFAR using the \texttt{dreamBeam}\footnote{\url{https://github.com/2baOrNot2ba/dreamBeam}} package to mitigate this effect. While most bursts remained mostly unaffected by this correction (owing to high elevation of the source during detection by the telescope), it reduced the fraction of linear polarization detected for a few of the bursts by up to 12\%. Owing to the millisecond-duration of our bursts, effects of changing projection effects for the dipole length as the source is tracked in the sky will be negligible. LOFAR real-time calibration corrects for cable and geometric delays during beam-forming. We use \texttt{ionFR}\footnote{\url{https://github.com/csobey/ionFR}} \citep{Sotomayor-Beltran_2013_A&A} to calculate the ionospheric contribution to the measured RM (RM$_{\mathrm{iono}}$ in Table~\ref{tab:prop_effects}).

For polarization analysis, the complex voltage data was coherently dedispersed to a DM$ = 348.772$\,\pcpercc\ using \texttt{dspsr} \citep{vanstraten_2011_pasa} at a time resolution of $983.04$\,$\upmu$s and frequency resolution of $12.2$\,kHz. At this frequency resolution, depolarization due to uncorrected intra-channel Faraday rotation at the lowest-observed frequency in the band (109.88\,MHz) is less than 1\% at the maximum absolute RM value reported for \rthree\ by \citetalias{pleunisetal} (and $<12$\% up to an RM$=1000$\,\radpermsq). The linear hands give us the correlated polarization products: XX, XY, YX and YY that are used to calculate the 4 Stokes parameters $I$, $Q$, $U$ and $V$, with $I^{2} = Q^{2} + U^{2} + V^{2}$. Linear polarization is computed as $L_\mathrm{meas} = \sqrt{Q^{2}\;+\;U^{2}}$. The true linear polarization is obtained after de-biasing the linear polarization $L_\mathrm{meas}$ based on the prescription in \citet{everett_2001_apj}, where:

\begin{align}
L_\mathrm{true} = 
\begin{cases}
    \sqrt{ \left( \dfrac{L_\mathrm{meas}}{\sigma_{I}} \right)^2 - \sigma_{I}}  & \text{if \,} \dfrac{L_\mathrm{meas}}{\sigma_{I}} \geq 1.57 \\
    0, & \text{otherwise}
\end{cases}
\end{align}

Faraday rotation occurs as the electromagnetic wave passes through a magnetized plasma. It has the effect of rotating the plane of polarization of a linearly polarized wave, with the degree of rotation scaling as the inverse square of the radio frequency. For a cold, magnetized plasma, the change in PPA due to Faraday rotation ($\Psi$) can be expressed as a function of the wavelength ($\lambda$) as follows:
\begin{align}
    \Delta\Psi(\lambda) = \mathrm{RM}\; \lambda^{2} 
\end{align}
 where PPA $\Psi = \tfrac{1}{2} \tan^{-1}\frac{U}{Q}$ (periodic as $\pi$ in phase); and RM quantifies the slope of the PPA versus $\lambda^{2}$ relation.

We measure RM by implementing an RM-synthesis \citep{Burn_1966_MNRAS, Brentjens_2005_A&A} algorithm. The values obtained using our code were verified using the \texttt{RM-tools} software\footnote{\url{https://github.com/CIRADA-Tools/RM-Tools}} \citep{Purcell_2020_ascl}. \citet{Burn_1966_MNRAS} define the Faraday Dispersion Function (FDF) as the Fourier transform of the complex linear polarization vector per unit RM. The measured linear polarization vector is modified/multiplied by the `RM Spread Function' (RMSF) as a result of the finite window of bandwidth of the telescope. This window determines the resolution obtainable in RM-space, and is given by the FWHM of the RMSF. For the LOFAR setup used in this work, the RM resolution is $\approx \tfrac{3.8}{\lambda_\mathrm{max}^2-\lambda_\mathrm{min}^2} = 0.77$\,\radpermsq. In the case of FRBs, a Faraday-thin FDF is expected to first order, since we expect a near-perfect point source behind a Faraday screen \citep{michilli_2018_natur}. In this scenario, all the polarized flux is Faraday rotated by the same value --- i.e., the FDF can be approximated to a delta function at the value of a single RM, convolved with the RMSF. We obtain the RM value from the single peak of the Faraday-thin FDF. By definition, Faraday rotation affects Stokes $Q$ and $U$, being seen as oscillations in the $Q$ and $U$ spectra due to the rotation in the complex linear polarization vector ($Q \;+\; iU$) as a function of $\lambda^{2}$. We obtain the total linear polarization and the PPA of the pre-Faraday-rotated burst by `derotating' the $Q$ and $U$ spectra (integrated over the burst duration) by the determined RM as shown below:
 \begin{align}
     \big( {Q} \;+\; i{U} \big)_{derot} (\lambda) = \big( {Q} \;+\; i{U} \big)_{rot} (\lambda) \; e^{2i\mathrm{RM}\lambda^{2} - \Psi_{0} }  
 \end{align}
 where $\Psi_{0}$ is the PPA at infinite frequency.

 RM measurement was performed using only the corresponding frequency channels of the $Q$ and $U$ spectra occupied by the Stokes $I$ spectrum of the burst in order to increase S/N. The lack of polarization in B26 and B29 could in part be due to their very low S/N ($<8$). For B24 and B25 (note that B25 is at the top of the LOFAR band), we obtain no RM measurement despite higher S/N and lower scattering. The two visible components in B25's dynamic spectrum were also searched separately for a peak in the FDF. For these bursts that appear completely depolarized in the LOFAR band, we place conservative upper limits on $L/I$ after derotating the $Q$ and $U$ spectra by an `expected' RM value using the best-fit slope of 0.197 \radpermsq\ day$^{-1}$ from \citet{Mckinven_2023_ApJ_R3}, during their observed secular RM trend (since the true RM, assuming the intrinsic emission was polarized, is not known in these cases).

\citetalias{pleunisetal} reported on the polarimetric properties of 3 out of the 18 bursts they presented, since 14 of the bursts (B5 $-$ B18) had only total-intensity data available. In this work, we detect linear polarization from 4 out of 11 bursts and measure their RMs (bursts B19, B20, B23, and B27). We detect no significant circular polarization in any of these bursts. The remaining 7 bursts appear unpolarized. Polarization profiles of linear polarization, $L$, and circular polarization, Stokes~$V$, are plotted in the mid-panel of each sub-figure in Figure~\ref{fig:burstfamily} along with their total-intensity (Stokes~$I$) profile. The polarization profiles shown in Figure~\ref{fig:burstfamily}, like the total-intensity profiles, were only summed over the frequencies where the burst is present in the band. In the top panels for each burst are the corresponding PPAs. 

\begin{figure*}
  \centering
  \includegraphics[width=0.99\textwidth]{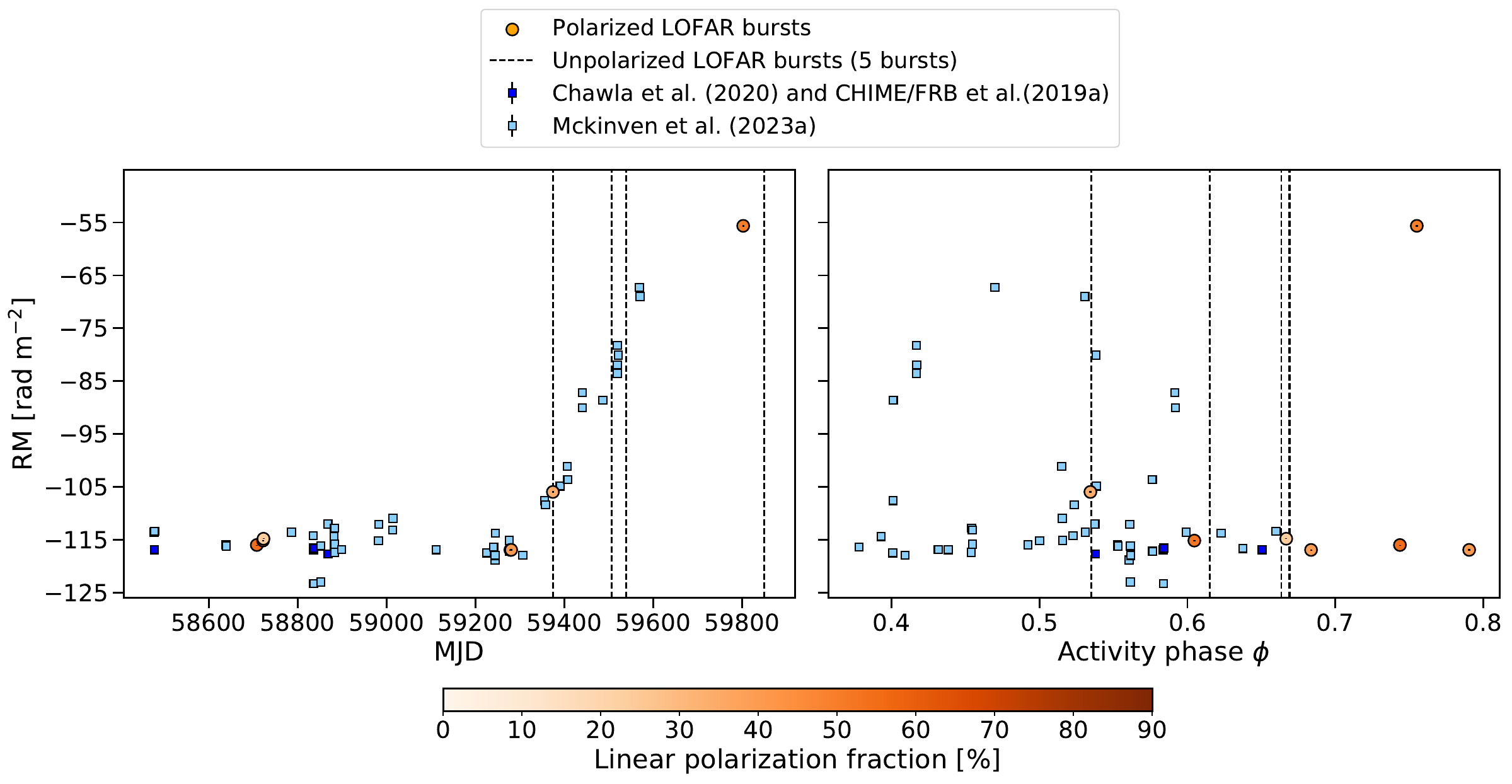}
  \caption{RM versus time ({\it Left}) and versus activity phase ({\it Right}). Until B19, the RM variations were only $\sim2-3$\,\radpermsq. The LOFAR RM measurements at $110-188$\,MHz agree with the RM trend observed by \citet{Mckinven_2023_ApJ_R3} at $400-800$\,MHz using CHIME/FRB. The LOFAR data points are also coloured by their corresponding linear polarization fractions, while unpolarized bursts are denoted by vertical dashed black lines at their respective arrival times and phases. Note that the linear polarization fractions vary with the central frequency of the bursts and with time (see Figure~\ref{fig:depol}).}
  \label{fig:RMs}
\end{figure*}

The measured RMs, assuming a $\lambda^{2}$ scaling of the PPAs (and the corresponding ionospheric contribution to the RM, RM$_\mathrm{iono}$), along with the linear polarization fraction measured at this RM can be found in Table~\ref{tab:prop_effects}. We plot the measured RMs as a function of time and activity phase in Figure~\ref{fig:RMs}. Until B22 on 2021 March 07, bursts from \rthree\ showed subtle but detectable RM variations in the LOFAR band of $\sim2-3$\,\radpermsq. For burst B23, we measure an RM $= -105.6$\,\radpermsq, i.e., an absolute change of $10.9$\,\radpermsq\ and a $\sim8$\% fractional change in the RM compared to the previous LOFAR measurement from B20. When over-plotted with RM measurements at $400-800$\,MHz from \citet{Mckinven_2023_ApJ_R3}, this apparently abrupt change in RM agrees with the secular trend in RMs observed in bursts detected with CHIME/FRB. The most recent RM measurement with LOFAR for burst B27, with RM$=-54.4$\,\radpermsq, points towards a decrease in the RM gradient with time.

\begin{figure*}
  \centering
  \includegraphics[width=0.9\textwidth]{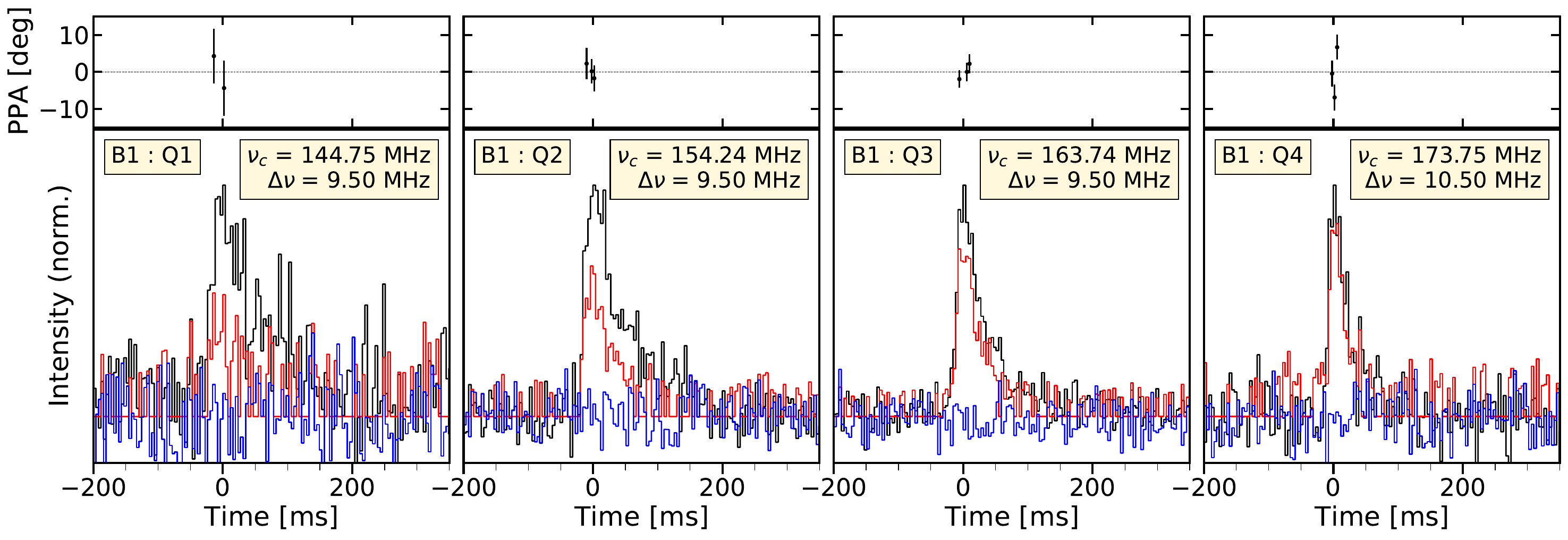} \\
  \vspace{0.7cm}
  \includegraphics[width=0.95\textwidth]{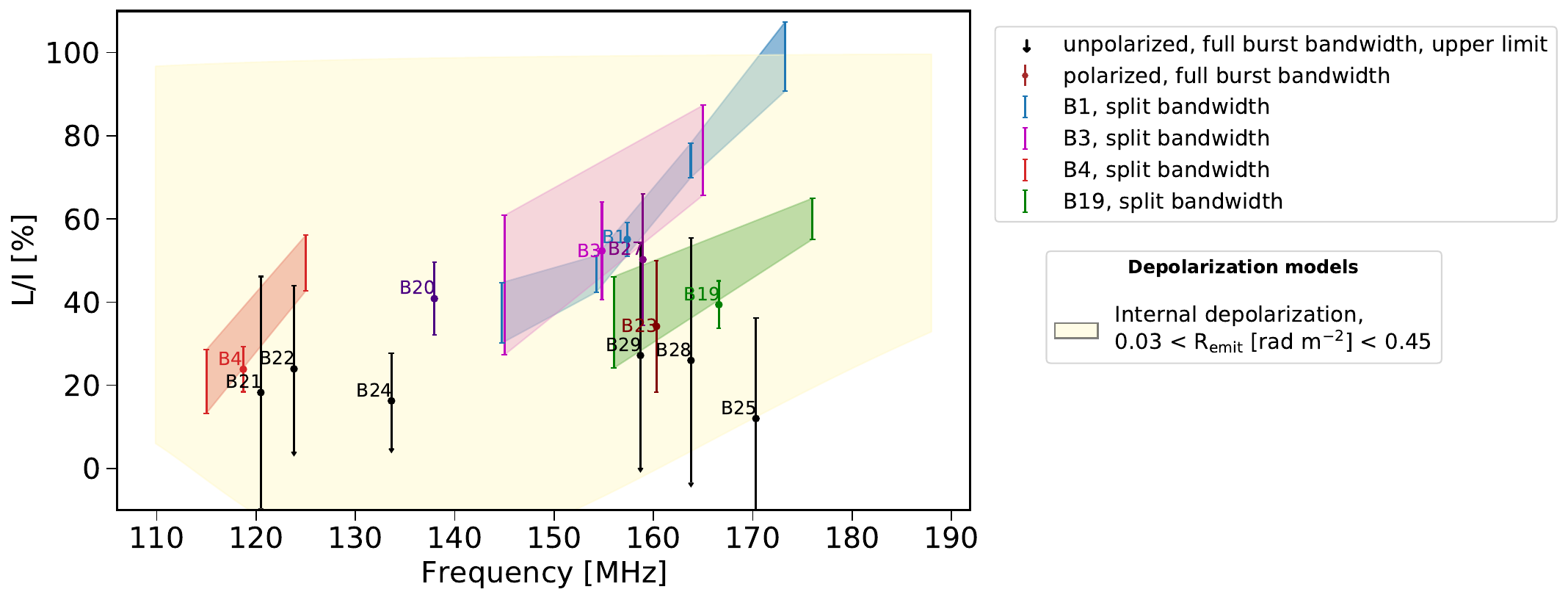}
  \caption{{\it Top}: Total intensity (black) and polarization profiles (linear: red; circular: blue) for a single burst (B1), at different radio-frequency ranges of the total spectral range occupied by the burst. The band occupied by the burst was divided into four quarters labelled Q1, Q2, Q3, Q4 from the lowest to highest frequency, respectively. The profiles for each quarter have been normalized by the peak of the total intensity, for visual purposes. The respective linear polarization fraction measured in these quarters can be found in the figure below. {\it Bottom}: Linear polarization fractions versus the central frequencies of the bursts. Additionally, B1 was split in bandwidth into quarters, and B3, B4, B19 into halves; and the linear polarization fractions of these parts plotted separately. We see that the linear polarization decreases with decreasing frequency even within the spectral extent of a single burst. Overplotted is the internal depolarization model (see Section~\ref{ssec:implications_propeffects} and Equation~\ref{eq:internal_depol}). Additionally, we see that bursts B25, B28, and B29 are completely depolarized despite occupying the top-half of the frequency band. Upper limits are calculated for the bursts with no detected polarization (B21, B22, B24, B25, B28, and B29) at an interpolated RM value based on the best-fit slope for the RM secular trend from \citet{Mckinven_2023_ApJ_R3}.}
  \label{fig:depol}
\end{figure*}

For LOFAR bursts from \rthree\ that had sufficient S/N (including B1$-$B3, published in \citetalias{pleunisetal}), we were able to divide the bursts into halves/quarters in the frequency band they occupied, and measure the linear polarization fraction in each of these frequency chunks. An example of the Stokes~$I$, $L$, and Stokes~$V$ profiles for burst B1 in quarters of its total occupied bandwidth is shown in Figure~\ref{fig:depol} (top). This allows for quantifying the variation of polarization fraction with frequency within the same burst. Figure~\ref{fig:depol} (bottom) has the split-bandwidth linear polarization fraction measurements, as well as the full-bandwidth values plotted against central frequency. It shows that \rthree\ bursts exhibit depolarization towards lower frequencies not only between bursts as reported in \citetalias{pleunisetal}, but also within the individual bursts. It is also clear from Figure~\ref{fig:depol} (bottom) that the frequency of depolarization is not constant between bursts in the LOFAR band.

We further check for RM jumps in time across the pulse profile (as seen in some pulsar profiles in \citealt{Noutsos_2009_MNRAS}) but do not find any evidence for this. However, we do find small variations in RM for the same burst at different parts of its frequency band. Any additional dependence of RM on wavelength/frequency, when fit for $\mathrm{RM} \propto \lambda^{2}$, can be seen as a deviation from this canonical $\lambda^{2}$ scaling. We find hints of deviation from the expected $\lambda^{2}$ scaling for one of the bursts; further analysis of this will be explored in a future paper.

\section{Discussion}
\label{sec:disc}

We see that the frequency-dependent activity window of \rthree\ at LOFAR remains stable in time. With our improved constraints on the activity window at 110$-$188\,MHz, we find that the width of this window ($4.3^{+0.7}_{-0.2}$\,day) is likely smaller than the $5.2$-day window in the $400-800$\,MHz range of CHIME/FRB. This is unlike what \citet{pastormarazuela_2021_natur} find from observations at higher frequencies, where the activity window at the lower frequencies of the 600\,MHz CHIME/FRB band is wider than that at the higher frequencies of $1370$\,MHz with Apertif. \citet{Bethapudi_2022_arXiv} detected the highest-frequency bursts from \rthree\ to date, by extrapolating the chromatic activity to frequencies of $4-6$\,GHz to pick observing windows; however, more observations distributed across all activity phases are required to constrain the width of the window at these frequencies. Along with the new LOFAR window constraints we present here, this suggests that while the systematic delay in activity spans nearly 6 octaves in frequency, a systematic increase in activity width towards lower frequencies does not hold true for the entire frequency range of detections to date. 

\subsection{Implications of the observed propagation effects}
\label{ssec:implications_propeffects}

We find that the drifting rate of sub-bursts (the `sad trombone' effect) does not depend on the activity phase, which means its cause should be decoupled from the mechanism causing the 16.33-day periodicity. The drifting could be caused by intrinsic emission processes, like radius-to-frequency mapping of the emission region \citep{hessels_2019_apjl}.

\citet{marcote_2020_natur} find a scintillation bandwidth of $59\pm{13}$\,kHz at 1.7\,GHz, corresponding to a scattering timescale of 0.015\,ms at 1\,GHz, after correcting for plane-wave scattering of an extragalactic burst \citep{Ocker_2021_ApJ}. This value is found to be consistent with the expected line-of-sight scattering from the interstellar medium (ISM) in the Milky Way disk based on the NE2001 model \citep{cordes_2002_arxiv}. At 150\,MHz this spans a range of scattering timescales between 14\,ms (scattering index $\alpha = -3.6$) and 63\,ms (scattering index $\alpha = -4.4$). The total scattering observed will be a linear sum of the line-of-sight contributions:

\begin{align}
    \tau_{s} = \tau_{\mathrm{s,mw}} + \tau_{s,\mathrm{s,halo}} + \tau_{s,\mathrm{host}} + \tau_{s,\mathrm{frb}}
\end{align}

, where $\tau_{\mathrm{s,mw}}$, $\tau_{s,\mathrm{s,halo}}$, $\tau_{s,\mathrm{host}}$, $\tau_{s,\mathrm{frb}}$ are scattering contributions from the ISM of the Milky Way, the Milky Way halo, the ISM of the host, and from the circum-source medium of the FRB, respectively. $\tau_{s,\mathrm{s,halo}}$ is $< 4$\,ms at 150\,MHz based on the estimate from \citet{Ocker_2021_ApJ}. We find that $\tau_{\mathrm{s,mw}}$ and $\tau_{s,\mathrm{s,halo}}$ can account for nearly all of the scattering we measure in the bursts at 150\,MHz, with minimal contribution from $\tau_{s,\mathrm{host}}$. We find hints of a few ms-timescale variations on weeks to months timescales that could arise from the local environment of the FRB. We do not find strong evidence for longer timescale variations on years timescales, that could possibly arise from the relative motion of the observer and the Milky Way and host galaxy ISM. 

Up to factor-two variations in scattering time over a few minutes to days timescales have been reported for another repeating FRB, FRB\,20190520B \citep{Ocker_2023_MNRAS}. Such variations are also observed in the Crab pulsar, accompanied by DM variations, and are attributed to arise from the discrete structures (`filaments') of the Crab nebula \citep{McKee_2018_MNRAS, Driessen_2019_MNRAS}. 

Rotation measure variations ($>5$\% fractional), both drastic and moderate, have been previously seen in various other repeating FRBs: \rone\ \citep{michilli_2018_natur,hilmarsson_2021_apjl,Plavin_2022_MNRAS}, FRB\;20180301 \citep{luo_2020_natur}, \rsixseven\ \citep{hilmarsson_2021_mnras, Xu_2022_Natur, Jiang_2022_arXiv}, FRB\;20190520B \citep{Anna-Thomas_2022_arXiv}, and 12 additional repeating FRBs as shown in \citet{Mckinven_2023_ApJ_repeaterspol}. \rone\ and FRB\;20190520B exhibit extreme RM variations $>10^{4}$\,\radpermsq, and both have a compact persistent radio source associated with them \citep{chatterjee_2017_natur,marcote_2017_apjl,Niu_2022_Natur}. Out of these, the two periodically active repeaters \rone\ and \rthree\ reside in vastly different types of galaxies, while both being close to star-forming regions. Until MJD~59355, \rthree\ was in fact one of the only repeaters to exhibit relatively minor RM variations \citepalias[$2-3$\,\radpermsq;][]{pleunisetal} in comparison, despite these variations being higher than typically observed towards Galactic pulsars. While it is not possible to comment on the RM evolution of non-repeating sources, repeating FRBs seem to favour residing in dynamic magneto-ionic environments.
 
The phase centre and width of the activity window remain consistent between bursting activity observed before and after MJD~59355 (although more detections are required to obtain better constraints after MJD~59355), when the first burst with an RM change $>10$\,\radpermsq\ was detected by \citet{Mckinven_2023_ApJ_R3}. We do not detect any obvious change in the chromatic activity window at LOFAR accompanying the large RM variations following relatively stable RM measurements over $\sim$2.4\,years. Consistent values between CHIME/FRB and LOFAR RM measurements (including the latest RM measurement reported in this work, as conveyed to us by private correspondence with Ryan Mckinven) show that the RM variations are unlikely to be dependent on the frequency band in which the bursts are detected. This implies that the emission at different frequencies comes from the same Faraday depth, thus placing the Faraday screen(s) where the RM variations originate, in front of all the emission region(s) at different frequencies.

In the absence of any significant correlation between the small ($< 0.8$\,\pcpercc) DM variations with the RM variations, \citet{Mckinven_2023_ApJ_R3} attribute the observed RM evolution to changes in the strength\slash direction of the magnetic field in the local medium producing Faraday rotation (and not changes in the electron density of the Faraday screen). They calculate limits on the rate of change in the line-of-sight component of the magnetic field between a fraction of a $\mu\mathrm{G}/\mathrm{day}$ and $\sim 10^{5}$\,$\mu\mathrm{G}/\mathrm{day}$, depending on the assumed DM contribution from the Faraday-active medium (the greater the assumed DM contribution, the lower the required change in the magnetic field to account for a given change in RM). 

Bursts from \rthree\ typically have lower fractional linear polarization towards lower frequencies, even within individual bursts (Figure~\ref{fig:depol}). However, there is a large scatter in the  fractional polarization between different bursts, even at the same radio frequency (Figure~\ref{fig:depol}). For instance, consider bursts B1, B3, B19, B23, B27 which are centered within a 10-MHz range between 157$-$167\,MHz. Despite being close in emission frequency, these bursts show variations in linear polarization fractions in the range $\sim34$\% to $\sim60$\%. B25, B28 and B29, also centered at the top of the band show no detectable polarization at all. This is comparable to the range of polarization fractions observed across the entire 80-MHz LOFAR band.

\citet{beniamini_2022_mnras} discuss depolarization effects on FRBs caused by multi-path propagation through a magnetized scattering screen. They expect a a power-law dependence of the polarization fraction, decreasing with frequency. \citet{Feng_2022_Sci} found that a model of exponential decrease in polarization with frequency due to a scattering screen can explain the decrease in $L/I$ at lower frequencies for five repeating FRB sources. The analysis used by \citet{Feng_2022_Sci} applies a model originally developed by \citet{Burn_1966_MNRAS} and extended in a series of works since then \citep[e.g.,][]{Sokoloff_1998_MNRAS,Brentjens_2005_A&A}. These works focus on depolarization of incoherent synchrotron radiation. Their results, and in particular the exponential cutoff of the polarization below a certain frequency threshold differ qualitatively and quantitatively from the depolarization that will be suffered by coherent radiation passing through a scattering screen, as explored in \citet{beniamini_2022_mnras}. The latter is the appropriate case for FRBs, whose enormous brightness temperature implies that the radiation must be coherent.

It is possible that the large scatter in the polarization fractions of bursts centered around similar frequencies, when bursts from multiple activity cycles are considered, arise due to different lines-of-sight through a Faraday active scattering screen in the local environment of the source. There might still be more constrained trends within each activity cycle that get washed out when bursts from multiple activity cycles are considered. On the contrary, we do not find a significant correlation between the scattering timescales of the bursts (at the central frequency of the band occupied by the burst) and linear polarization fractions of the bursts (see Figure~\ref{fig:pol_vs_scattering}).

In a scattering model of depolarization, the depolarization is caused by an external Faraday screen separate from the emitting region. Depolarization can also occur in a Faraday-active medium co-spatial with the emitting region. The plane of polarization undergoes differential Faraday rotation in the emission region itself, causing depolarization when summed over the extended region. The fractional depolarization ($\mathrm{f}_{\mathrm{depol}}^{\mathrm{int}}$) due to this `internal depolarization' model \citep{Gardner_1966_ARA&A, Burn_1966_MNRAS, Sokoloff_1998_MNRAS} can be described as:
\begin{align}
    \mathrm{f}_{\mathrm{depol}}^{\mathrm{int}} = 1 - \frac{\mathrm{sin}(2\mathrm{R}_\mathrm{emit}\lambda^{2})}{(2\mathrm{R}_\mathrm{emit}\lambda^{2})}
\label{eq:internal_depol}
\end{align}
for the simple case of a uniform slab with a regular magnetic field and Faraday depth $\mathrm{R}_\mathrm{emit}$, where the differential Faraday rotation $\Delta \theta$ due to the Faraday depth of the slab is $\mathrm{R}_\mathrm{emit}\lambda^{2}$. By eye, we find that $\mathrm{R}_\mathrm{emit}$ of the emitting region would occupy a range between $\sim$0.03 and $\sim$0.45\,\radpermsq\ (Figure~\ref{fig:depol}). The best-fit value of 0.18\,\radpermsq\ at the 600-MHz central frequency of the CHIME/FRB band \citep{Mckinven_2023_ApJ_R3} also lies within this range. Note that, in this case, no correlation between the temporal scattering (caused by external foreground screens) and depolarization fraction is expected, which agrees with our findings (Figure~\ref{fig:pol_vs_scattering}). \citet{Plavin_2022_MNRAS} found that internal depolarization by an emitting region of Faraday depth 150\,\radpermsq\ could explain the depolarization seen for bursts from the possibly periodically active, repeating source \rone\ from 1.7$-$5\,GHz. Given that the emission at 150\,MHz and 600\,MHz are likely from different emission regions (since they are band-limited and happen at different activity phases), the RM evolution is likely still caused by a different Faraday screen along our line-of-sight to the different emission regions. 

\subsection{A self-consistent model for the observed propagation effects}

\rthree\ is periodically active for a total 8.8-day window every $\sim$16.3\,days, first at higher frequencies and a few days later at lower frequencies (from 6\,GHz down to 110\,MHz). Explanations for the periodic activity of \rthree\ have so far come in three classes: a binary orbit, precession, or the rotation of a compact object. We will briefly review these three hypotheses and discuss them in context with observational constraints on the source.

\textit{Binary orbit:} \citet{tendulkar_2021_apjl} favour a high-mass X-ray binary model (with a late O-type or B-type companion) given \rthree's 16-day periodic activity and inferred age of $\mathcal{O}({10^{6}} \mathrm{yr})$ based on the observed offset from a star-forming knot in its host galaxy. They suggest that the modulation of activity window with frequency in such a scenario would come from free-free absorption in the swept back wind of the companion \citep{ioka_2020_apjl, wada_2021_apj}. Other proposed binary models include a Be/X-ray binary system \citep{li_2021_apjl} with a NS and Be-star, and an eccentric binary system consisting of a magnetar and early-type star \citep{Barkov_2022_arXiv}. In \citet{li_2021_apjl}, interactions between the circumstellar disk of a companion Be-star and the NS produce starquakes on the NS resulting in FRBs, while free-free absorption in the disk causes a chromatic activity window. If the RM evolution is the result of the periastron passage of the NS through such a disk, then the orbital period must be greater than the 3-year timescale of RM evolution observed for the source thus far. Such a system would need an alternate explanation for the observed 16.3-day activity period. A well-known example of a similar system is the Galactic binary containing PSR~B1259$-$63 and a Be star. In the 50 days around the periastron in its 1237-day orbit, changes in DM ($\sim$ 6$-$8\,\pcpercc), RM ($\sim$6000\,\radpermsq), and an increase in scattering (such that the pulsar gets eclipsed by the companion disk for 35 days) were observed \citep{Johnston_1992_ApJL, Johnston_1996_MNRAS, Johnston_2005_MNRAS}. These drastic changes in the propagation effects undergone by the pulses are thought to arise from the stellar disk of the Be-star companion. Alternatively, an eccentric binary system where the emission frequency depends on the orbital separation, along with viewing angle limitations from beaming geometry, could explain the asymmetric chromatic activity windows seen for \rthree.

\textit{Slow rotation:} \citet{beniamini_2020_mnras_496} proposed that \rthree\ is an ultra-long-period magnetar that rotates every $\sim$16.3 days. In this scenario, the emission is associated with a polar cap beam with a high-to-low-frequency chromaticity from radius-to-frequency mapping \citep[see also the simulations in][]{li_2021_apjl}. Any DM variations should not vary periodically with the rotational phase. The recent discoveries of radio emission from sources with periodicities of 76\,s \citep{Caleb_2022_NatAs} and 1091\,s \citep{Hurley-Walker_2022_Natur} and a 6.7-h magnetar candidate \citep{DeLuca_2006_Sci} add credence to the existence of a population of ultra-long-period magnetars, that most surveys are biased against detecting \citep{Beniamini_2023_MNRAS}. However, a 2 week period is still a stretch from the minutes--hours rotational periods that have so far been observed to exist.

\textit{Precession:} Precession of the FRB-emitting NS\slash magnetar may arise either from a freely precessing, non-spherical star deformed by its internal magnetic field \citep{levin_2020_apjl, zanazzi_2020_apjl}; orbitally induced forced precession of the NS in a NS - black hole (BH) binary \citep{yang_2020_apjl}; forced precession produced by a fall-back disk \citep{Tong_2020_RAA}; or, a tidal-force-induced precession of an emitting jet \citep{Chen_2021_ApJ_jetprecession}. \citet{li_2021_apjl} show that the chromatic windowing in the activity of \rthree\ can be reconciled with a slowly rotating or freely precessing magnetar with a curvature radiation emission model (with radius-to-frequency mapping), with asymmetric emission about the magnetic dipole axis. The forced precession scenarios require the activity windows at all frequencies to be centered around the same precessional phase, rendering them implausible. Precession models can be tested by monitoring for a time derivative of the activity period \citep[which would also be the precessional period;][]{Wei_2022_A&A}. Our observed flat PPAs within each burst are consistent with measurements in the same frequency band in the past \citepalias{pleunisetal} as well as at higher frequencies \citep{nimmo_2021_natas, pastormarazuela_2021_natur}, which are accommodated in both free and forced precession models by \citet{li_2021_apjl}. We do not calculate the absolute value of the PPAs, which if compared with those at higher frequencies would further constrain precession and slow rotation models that predict an evolution of the PPAs with activity phase.

The activity of the \rthree\ in its 16.33-day periodic window is frequency dependent, and we see in this work that the chromaticity remains stable with time in the LOFAR band (Section~\ref{ssec:activity}). However, this chromaticity does not seem to extend to the RM variations, with lower frequency LOFAR measurements of RM (of bursts B23 and B28) agreeing with the RM trend observed at higher frequencies by CHIME/FRB. In the context of binary models where the activity is modulated by the stellar wind or the disk of a companion, we infer that the RM variations must arise from a medium distinct from the one responsible for the chromatic activity windows. This strengthens the same conclusion drawn by \citet{Mckinven_2023_ApJ_R3} based on the fact that the RM variations occur on much longer timescales than the 16.3-day periodic activity. Such a magnetized screen that affects bursts occurring at different parts of the activity phase must exist between the source and us along our line-of-sight. This screen can be conceived to be a supernova remnant or a pulsar wind nebula (like the Crab nebula), although observations of the field around \rthree\ do not detect any persistent radio counterpart on milliarcsecond to arcsecond scales \citep{marcote_2020_natur}. In case of an expanding supernova remnant, the $|\mathrm{RM}|$ would decrease (mostly) monotonically over much longer timescales \citep{Yang_2023_MNRAS} than the source has been currently observed for.

\section{Conclusions}

\rthree, with its 16.33-day periodic activity is one of the most prolific repeating FRBs. This allowed us to track its activity, as well as the observable properties of the bursts in the $110-188$\,MHz LOFAR band with high cadence, over years timescales. We have tracked frequency-dependent variations in the spectro-temporal and polarimetric properties of the burst, some of them caused by propagation effects in the intervening medium, and find no correlations with the chromatic activity of the source. Continued monitoring of the source with LOFAR and other telescopes will chart the trajectory of the observed RM evolution, which transitioned from being stochastic to a secular decrease. This will provide strong constraints on any self-consistent model of this prolific repeating FRB source.

\section*{Acknowledgements}

We thank the anonymous referee for helpful comments that improved the quality of the manuscript. We also thank Caterina Tiburzi and Krishnakumar Moochickal Ambalappat for their help with the calibration of projection effects in the LOFAR data, which impacts our polarization analysis. We thank Paz Beniamini for insightful discussions about depolarization.

Research by the AstroFlash group at University of Amsterdam, ASTRON and JIVE is supported in part by an NWO Vici grant (PI Hessels; VI.C.192.045).

This paper is based on data obtained with the International LOFAR Telescope (ILT) under project codes LC15\_014 and LT16\_006, in addition to ILT data used in \citetalias{pleunisetal}. LOFAR (van Haarlem et al. 2013) is the Low Frequency Array designed and constructed by ASTRON. It has observing, data processing, and data storage facilities in several countries, that are owned by various parties (each with their own funding sources), and that are collectively operated by the ILT foundation under a joint scientific policy. The ILT resources have benefited from the following recent major funding sources: CNRS-INSU, Observatoire de Paris and Universit\'{e} d'Orl\'{e}ans, France; BMBF, MIWF-NRW, MPG, Germany; Science Foundation Ireland (SFI), Department of Business, Enterprise and Innovation (DBEI), Ireland; NWO, The Netherlands; The Science and Technology Facilities Council, UK.

Z.~P. is a Dunlap Fellow. The Dunlap Institute is funded through an endowment established by the David Dunlap family and the University of Toronto.
\section*{Data Availability}

The raw complex voltage data of the observations of \rthree\ with LOFAR are available on the LOFAR Long Term Archive\footnote{https://lta.lofar.eu/} under project codes LC15\_014 and LT16\_006. The relevant code and data products for this work will be uploaded on Zenodo at the time of acceptance.
 



\bibliographystyle{mnras}
\bibliography{references} 




\appendix

\section{Additional figures}
\label{sec:appendix_fig}
{
\setcounter{figure}{0}
\renewcommand{\thefigure}{A\arabic{figure}}

\begin{figure*} 
  \centering
  \includegraphics[width=0.65\textwidth]{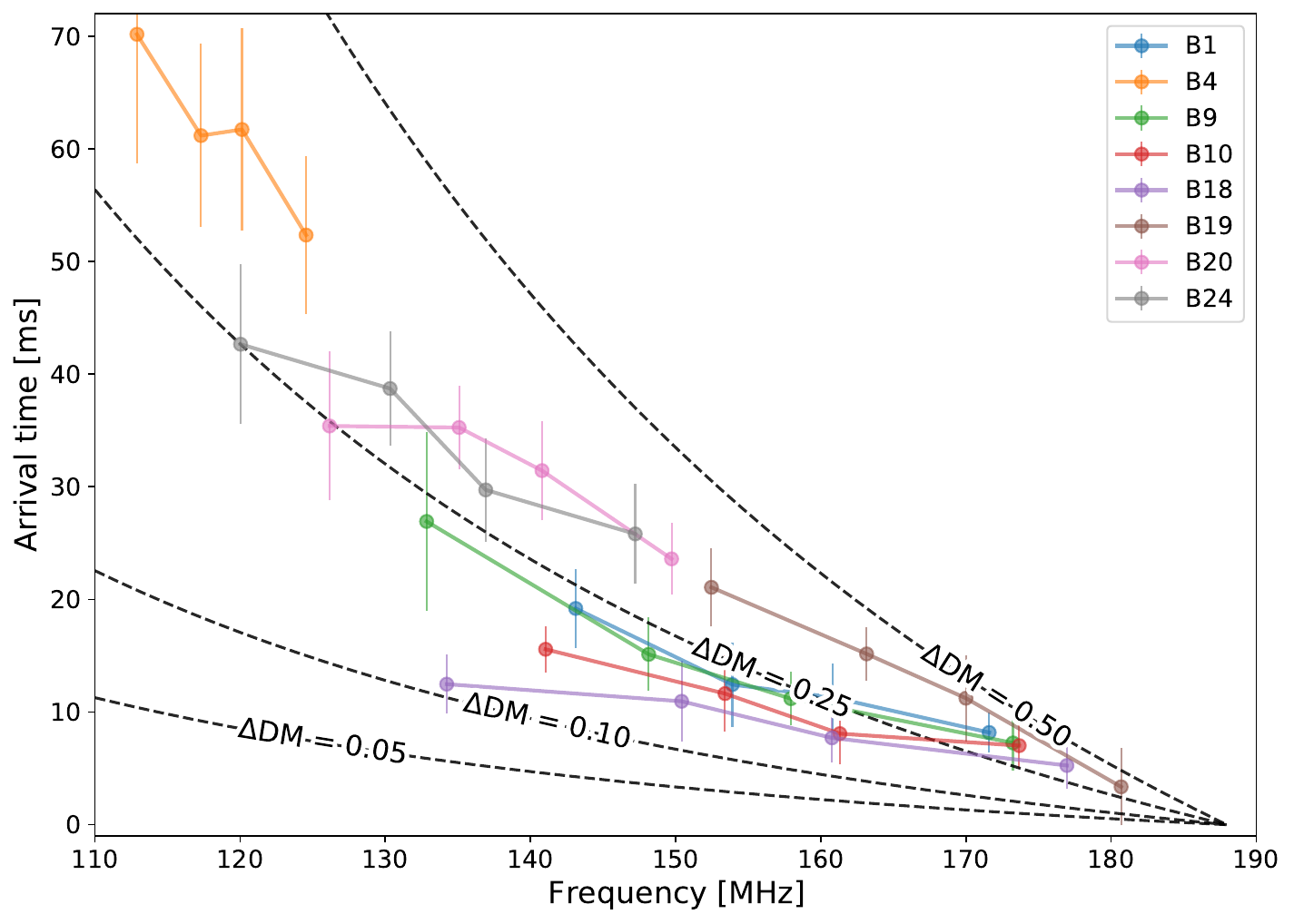}
  \caption{Times-of-arrival of the highest-S/N individual bursts at four equipartitioned subbands occupied by the burst. The peak of a Gaussian, convolved with an exponential tail, and fit to the subband integrated profile of the burst is taken as the arrival time. All the bursts are dedispersed to a DM$= 348.772$\,\pcpercc\ \citep[][see the main text for detailed discussion]{nimmo_2021_natas}. The $\Delta \mathrm{DM}$ curves for possible uncorrected positive DM deviations of 0.05, 0.1, 0.25 and 0.5\,\pcpercc, as referenced to the top of the LOFAR band, are plotted as black dashed curves. A least-squares fit to the slopes of the arrival time-frequency data points is used as a second method to calculate the drift rates of the bursts, the values of which are in Tables~\ref{tab:prop_effects} and \ref{tab:B1_to_B18_burst_properties} as well as Figure~\ref{fig:driftrate}.The delays being larger for bursts at the bottom of the band can be explained by the expected linear scaling of drift rate with frequency. }
  \label{fig:toa_vs_freq}
\end{figure*}

\begin{figure*} 
  \centering
  \includegraphics[width=0.9\textwidth]{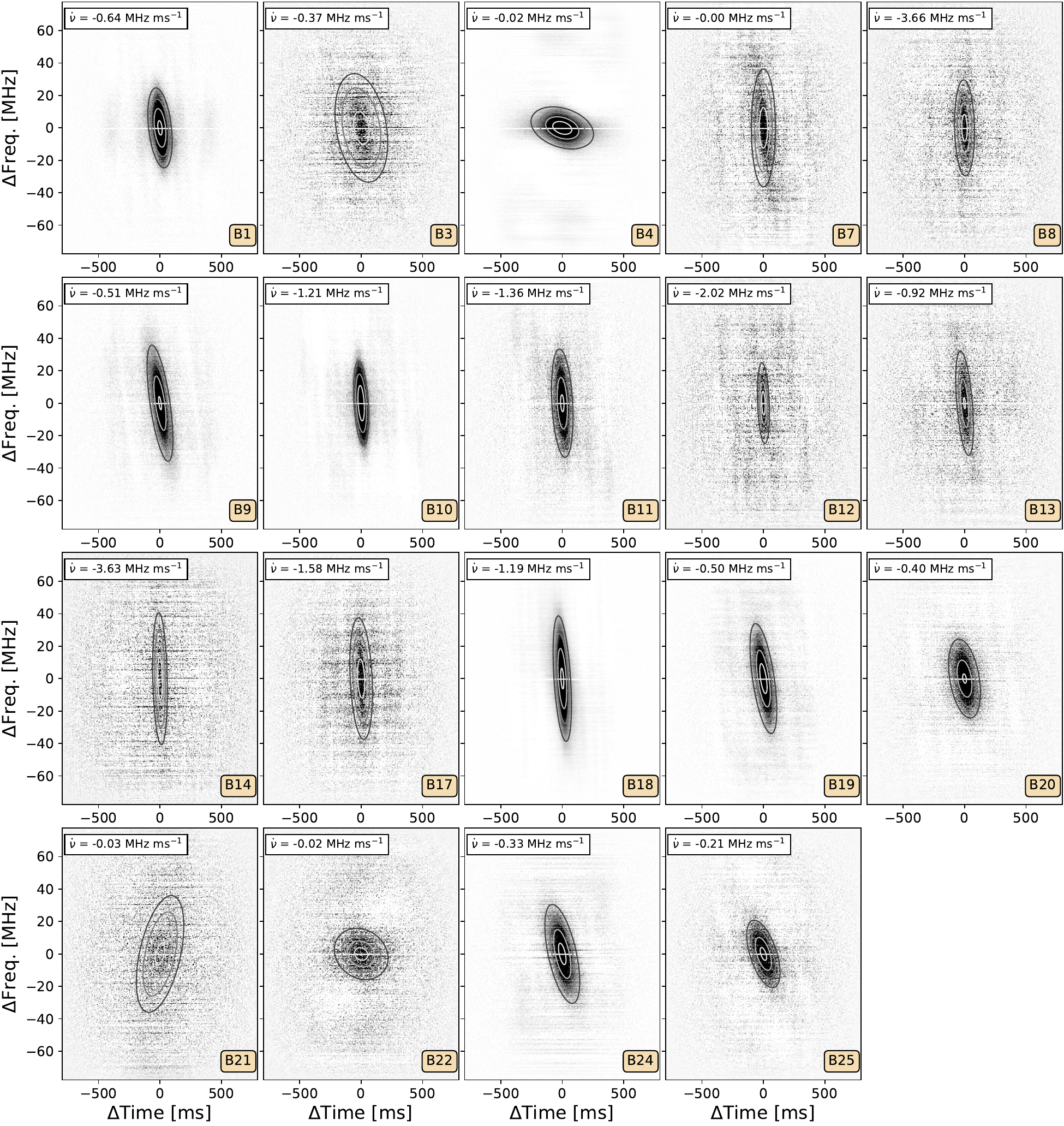}
  \caption{The 2D ACFs of the dynamic spectra of bursts B1$-$B29 (with S/N $> 10$) that are used to calculate the drift rates. All the dynamic spectra were dedispersed to a DM$=348.772$\,\pcpercc\ before computing the ACF \citep[][see the main text for detailed discussion]{nimmo_2021_natas}. The 2D tilted Gaussian ellipsoid fits to the ACFs are depicted as gray contours (showing the $1\sigma$, $2\sigma$, and $3\sigma$ levels).}
  \label{fig:drifts_familyplot}
\end{figure*}

\begin{figure*} 
  \centering
  \includegraphics[width=0.65\textwidth]{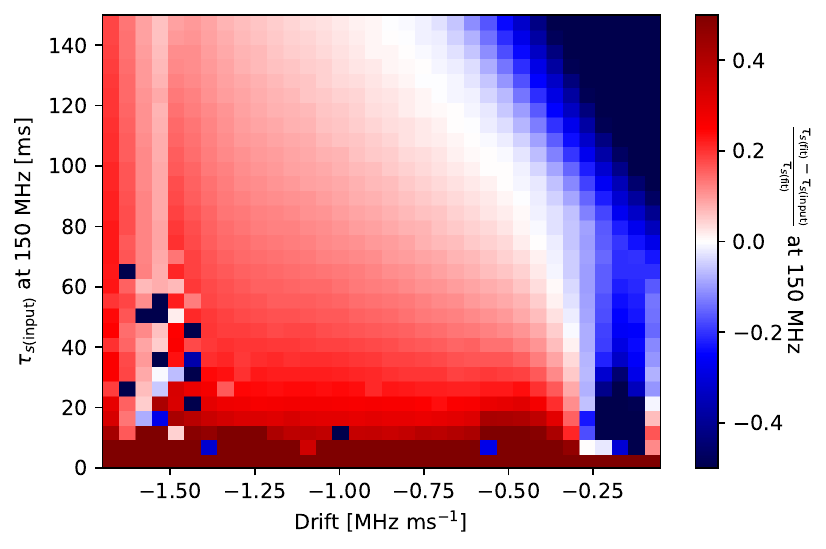}
  \caption{Difference between the measured scattering timescale at $150$\,MHz and the input scattering timescale, for simulated bursts with varying input drift rates and scattering timescales. See Section~\ref{ssec:scatteringdrift} for a full description of the simulated bursts. This is used to place systematic uncertainties on the scattering timescale fits for the detected bursts, accounting for covariance with the drift rate of the burst.}
  \label{fig:driftscatteringfit}
\end{figure*}

\begin{figure*} 
  \centering
  \includegraphics[width=0.75\textwidth]{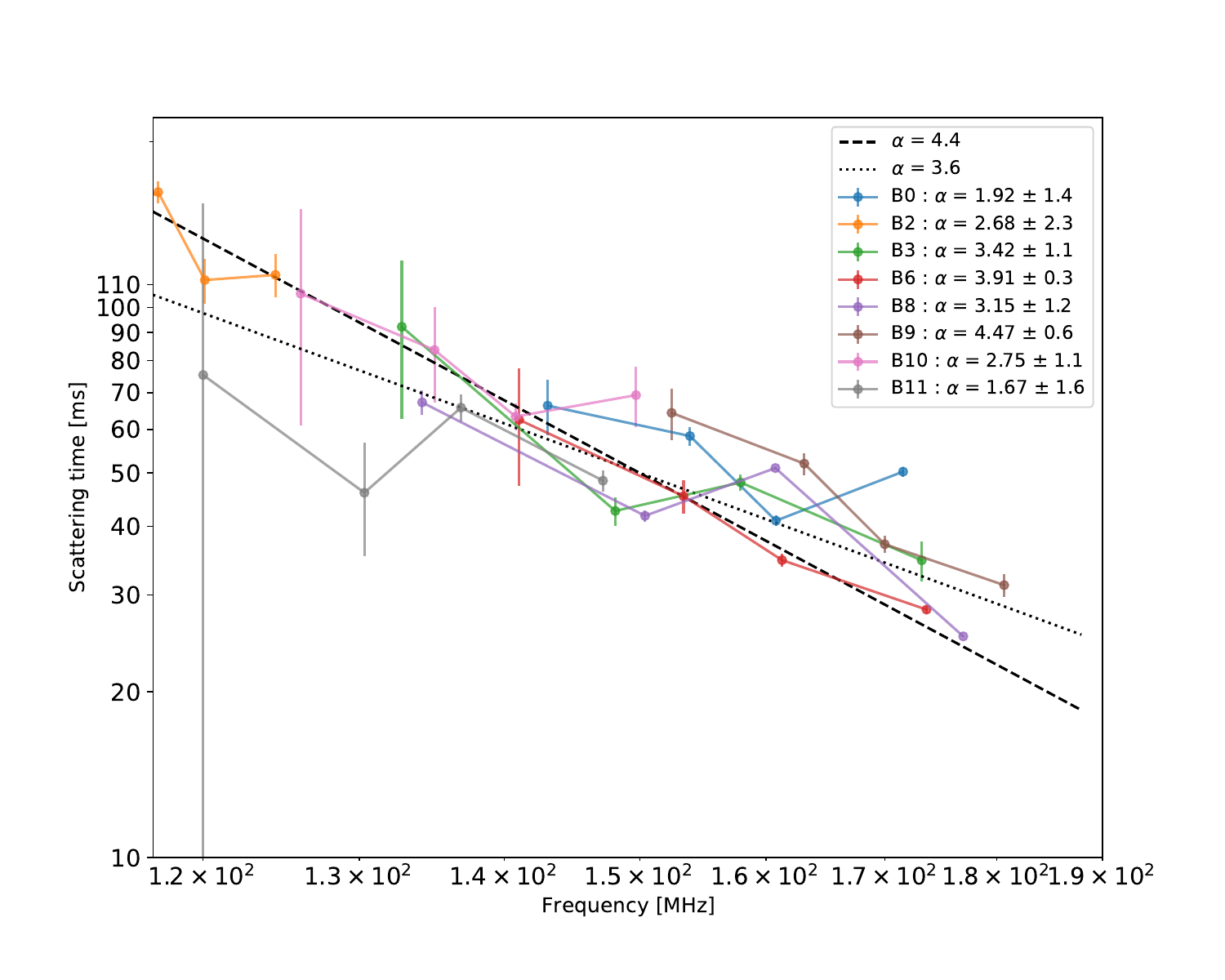}
  \caption{Scattering timescales measured at four equipartitioned subbands for each burst, versus frequency. Both the axes are logarithmic. The fitted scattering indices, $\alpha$ (where scattering scales with frequency as $\tau_\mathrm{s} \propto \nu^{\alpha}$), for each burst are shown in the legend along with the 1-$\sigma$ errors.}
  \label{fig:scattering_index_fit}
\end{figure*}

\begin{figure*} 
  \centering
  \includegraphics[width=\textwidth]{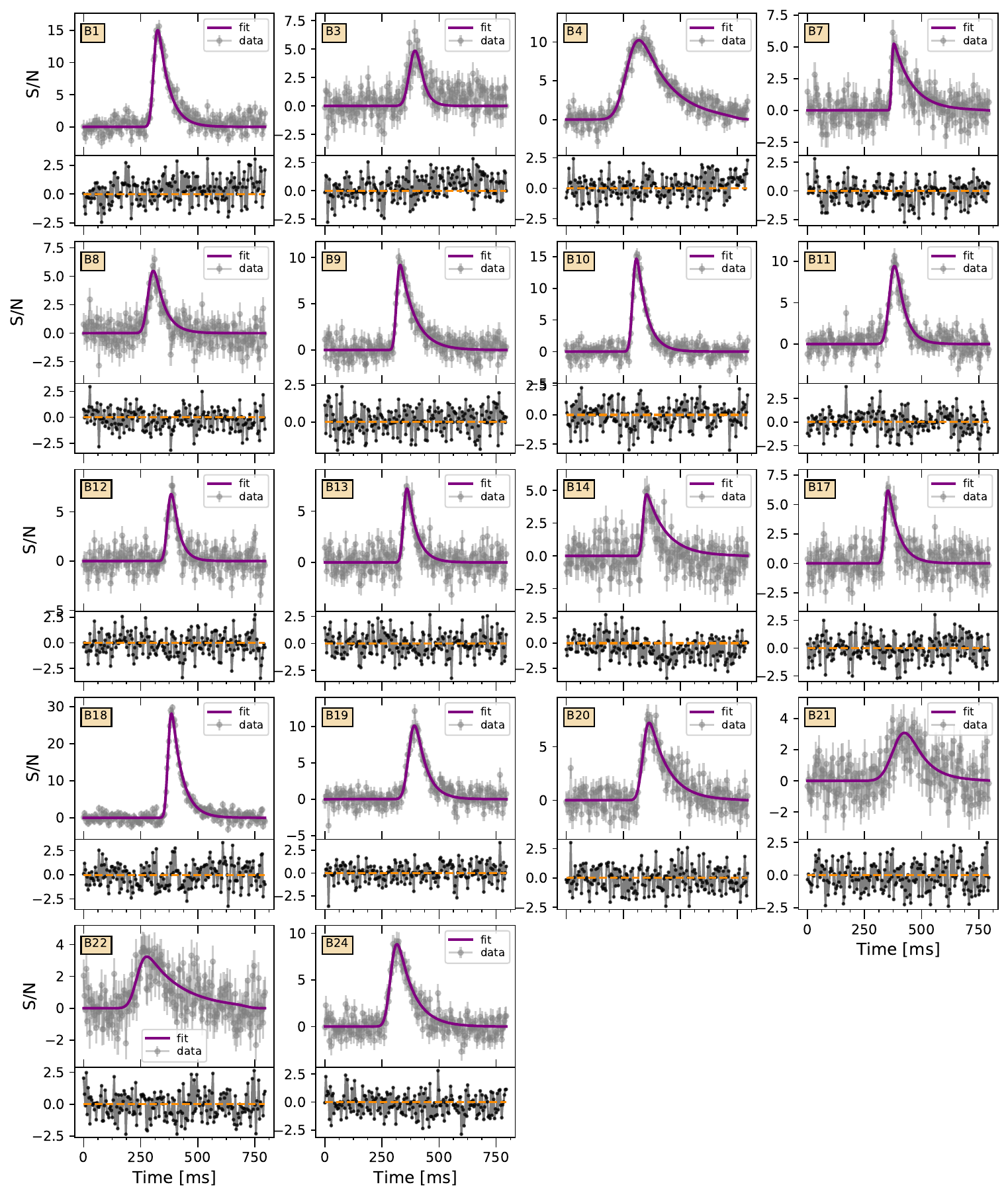}
  \caption{\textit{Top panel of each sub-figure}: The MCMC-based fit of a Gaussian, convolved with an exponential tail model, to the frequency-integrated time profiles of bursts with S/N$>10$, as described in Section~\ref{ssec:scatteringdrift}. The data is at a time resolution of $3.93$\,ms. \textit{Bottom panel of each sub-figure}: Difference between the fit model and measured burst profile in black (with the orange dotted line showing the line of zero difference). The fit for B14 might be overestimating the scattering timescale due to the presence of a smaller second peak in the time series profile of the burst, arising from a potential second burst component. Note that the bursts were dedispersed to a DM$= 348.772$\,\pcpercc\, see Section~\ref{ssec:scatteringdrift} in the main text for a detailed justification of why we used this single DM value.}
  \label{fig:scattering_residuals}
\end{figure*}

\begin{figure*} 
  \centering
  \includegraphics[width=0.45\textwidth]{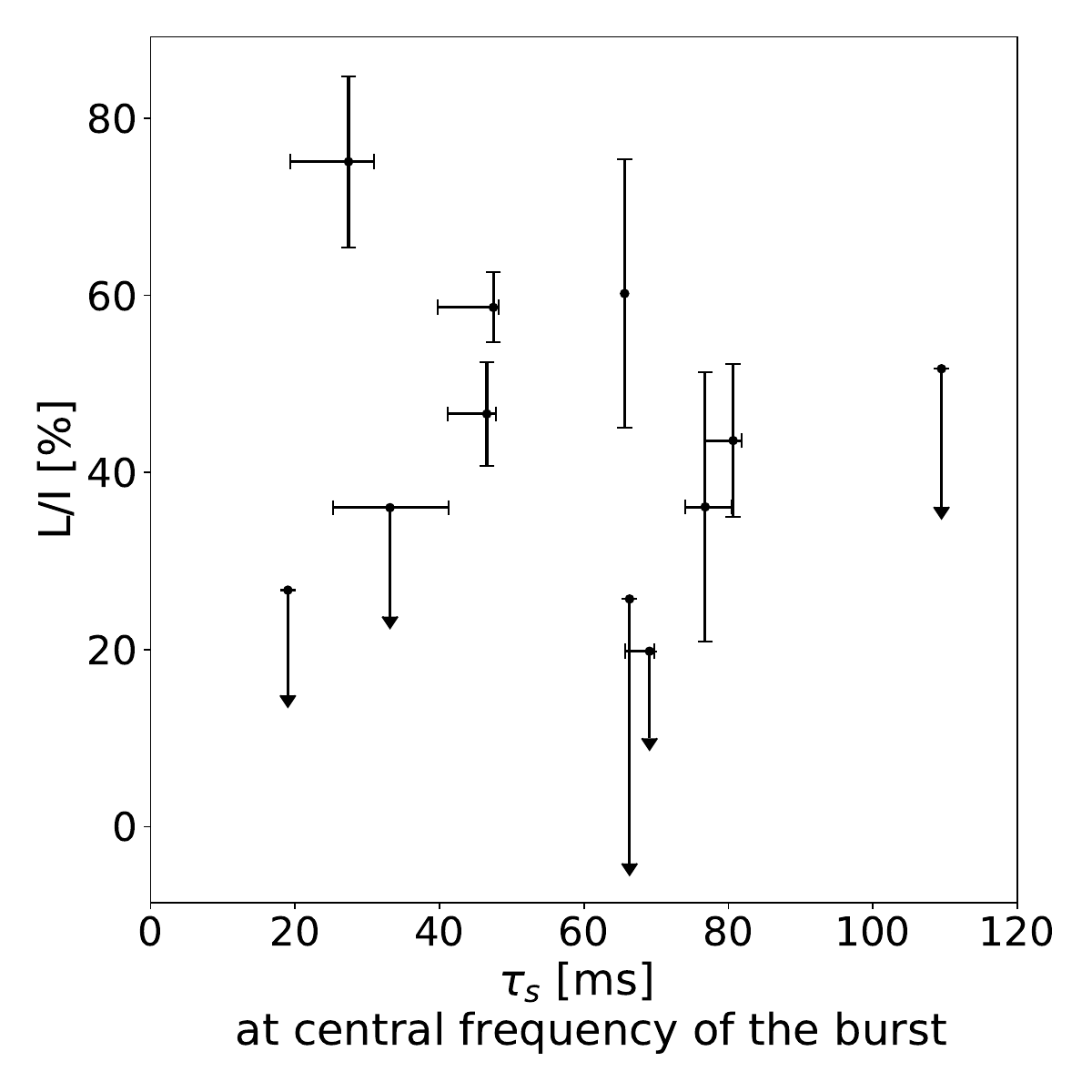}
  \caption{Linear polarization fraction versus the scattering timescales of bursts at 150\,MHz. The Pearson correlation coefficient between the two is $-0.21$, indicating a very weak correlation, if any. }
  \label{fig:pol_vs_scattering}
\end{figure*}

}

\section{Supplemental tables}

We recalculated the Gaussian burst widths and scattering timescales for bursts B1$-$B18 from \citetalias{pleunisetal} based on the MCMC fitting method described in Section~\ref{ssec:scatteringdrift}, and additionally calculate their drift rates. They are presented in Table~\ref{tab:B1_to_B18_burst_properties} (see Tables~\ref{tab:B19_to_B29_burst_properties} and \ref{tab:prop_effects} for B19$-$B29). These measurements are included in the figures and results in the main text. We find that the Gaussian burst widths calculated in this work (Table~\ref{tab:B1_to_B18_burst_properties}) are lower than the ones reported in Table~1 of \citetalias{pleunisetal} for the same bursts. We attribute this to the fitting models being different. \citetalias{pleunisetal} fit a Gaussian function to the time series of the bursts without fitting for the scattering tail for most of the bursts, while we fit a Gaussian convolved with an exponential tail to account for scattering.
\label{sec:appendix_table}
{
\begin{table*}
  \centering
  \footnotesize
  \caption{Measurements of propagation effects for bursts previously published in \citetalias{pleunisetal}.} 
  \begin{tabular}{lcccccc}
    \hline
Burst & Gaussian burst width & {$\tau_s$ (at 150\,MHz)}$\mathrm{^{a}}$ & Drift rate (from ACF)$\mathrm{^{b}}$ & Drift rate (from time of arrival)$\mathrm{^{c}}$ \\
     ID & (ms) & (ms) & (\mhzperms) & (\mhzperms) \\
    \hline
B01 &   30.5$\pm{0.6}$ &  57$_{ -7}^{  +2}$ & $-0.7\pm{0.3}$ & $-2.5\pm{0.5}$ \\
B02 &   82.5$\pm{0.5}$ &  --                & --  &   --     \\
B03 &   47.7$\pm{2.1}$ &  31$_{ -5}^{  +1}$  &  $-0.37\pm{1.1}$ & -- \\
B04 &   90.7$\pm{1.7}$ &  52$_{-10}^{  +9}$  &  $-0.02\pm{3.8}$ & $-0.64\pm{0.1}$  \\ 
B05 &   12.4$\pm{1.2}$ &  --                & --   &  --     \\
B06 &   23.7$\pm{3.4}$ &  --                 & --   &  --    \\
B07 &    6.9$\pm{0.6}$ & $<113_{- 4}^{+160}$ &  $-0.00\pm{1.6}$ & --  \\
B08 &   37.5$\pm{2.7}$ &  62$_{-15}^{  +4}$ &  $-3.66\pm{7.2}$  & -- \\
B09 &   26.8$\pm{0.6}$ &  69$_{ -4}^{  +1}$  &  $-0.51\pm{0.2}$ & $-1.9\pm{0.4}$\\ 
B10 &   26.0$\pm{0.4}$ &  53$_{-10}^{  +2}$  &  $-1.21\pm{0.2}$ & $-3.4\pm{0.7}$\\ 
B11 &   40.7$\pm{1.3}$ &  53$_{-10}^{  +3}$ &  $-1.36\pm{2.0}$ & $-1.2\pm{0.9}$\\ 
B12 &   26.9$\pm{1.0}$ &  48$_{-10}^{  +4}$  &  $-2.02\pm{0.2}$  & -- \\
B13 &   20.3$\pm{0.8}$ &  58$_{ -9}^{  +4}$  &  $-0.92\pm{1.7}$  & -- \\
B14 &   19.8$\pm{1.2}$ & $<101_{-23}^{  +4}$ &  $-3.63\pm{0.9}$  & -- \\
B15 &   63.8$\pm{0.5}$ &  --                 & --  &  --       \\
B16 &   19.8$\pm{1.2}$ &  --                & --  &   --     \\
B17 &   24.4$\pm{1.0}$ &  53$_{ -6}^{  +1}$  &  $-1.58\pm{0.9}$  & -- \\
B18 &   25.1$\pm{0.2}$ &  53$_{-10}^{  +2}$  &  $-1.19\pm{0.2}$ & $-5.4\pm{0.8}$\\
    \hline
    \multicolumn{5}{l}{$\mathrm{^{a}}$ Scattering timescales for bursts B07 and B14 are quoted as upper limits due to the fit}\\
    \multicolumn{5}{l}{\hspace{0.6em} being biased higher from a small secondary peak in their frequency integrated profiles (see Figure~\ref{fig:scattering_residuals}).}\\
     \multicolumn{5}{l}{$\mathrm{^{b}}$ Drift rates using the ACF method are only quoted for bursts with $\mathrm{S/N}>10$.}\\
   \multicolumn{5}{l}{$\mathrm{^{c}}$ Drift rates using the time of arrival method are only quoted for bursts with $\mathrm{S/N}>20$.} \\
    \multicolumn{5}{l}{\hspace{0.6em} The S/N limit here is higher than for the ACF method since we are more limited by S/N by dividing} \\
    \multicolumn{5}{l}{\hspace{0.6em} individual bursts into subbands to calculate the times of arrival at each subband.}\\
   \end{tabular}
   \label{tab:B1_to_B18_burst_properties}
\end{table*}

}


\bsp	
\label{lastpage}
\end{document}